\documentclass[journal]{IEEEtran} 

\usepackage{amsmath}
\usepackage{amssymb,amstext}
\usepackage[center]{subfigure}
\usepackage{lineno,hyperref}
\usepackage{xcolor}
\usepackage{extarrows}
\usepackage{algorithm}
\usepackage{algorithmic}
\usepackage{setspace}
\usepackage{bm}
\usepackage{graphicx}
\usepackage[justification=centering]{caption}
\usepackage{cite}
\usepackage{subfigure}
\usepackage{url}
\usepackage{multirow}
\usepackage{tabularx}
\usepackage{booktabs}
\usepackage{tikz}
\usepackage{amsfonts}
\usepackage{float}
\usepackage{array}
\usepackage{cite}
\usepackage{cases}
\usetikzlibrary{arrows,positioning,fit,shapes,calc,decorations.pathreplacing,matrix,patterns}

\newcommand{\larrow}{\overset{\scriptscriptstyle\leftarrow}}
\newcommand{\rarrow}{\overset{\scriptscriptstyle\rightarrow}}

\begin{document}
\title{Efficient Channel Estimation for RIS-Aided MIMO Communications with Unitary Approximate Message Passing}
\author{Yabo Guo, Peng Sun, Zhengdao Yuan, Chongwen Huang, Qinghua Guo, \IEEEmembership {Senior Member, IEEE}, Zhongyong Wang, and Chau Yuen, \IEEEmembership {Fellow, IEEE} 
    \thanks{The work of Y. Guo, P. Sun, Z. Yuan and Z. Wang was supported by National Natural Science Foundation of China (61901417) and Science and Technology Research Project of Henan Province (212102210173, 212102210556, 202102210313, 212102210566).
	(Corresponding authors: Qinghua Guo and Zhongyong Wang.)}
	\thanks{Y. Guo, P. Sun and Z. Wang are with the School of Information Engineering, Zhengzhou University, Zhengzhou 450002, China (e-mail: ieybguo@163.com, iepengsun@zzu.edu.cn,  zywangzzu@gmail.com).}
	\thanks{Z. Yuan is with the Artificial Intelligence Technology Engineering Research Center, Open University of Henan, and also with the School of Information Engineering, Zhengzhou University, Zhengzhou 450002, China (e-mail: yuan\_zhengdao@163.com).}
	\thanks{C. Huang is with the College of Information Science and Electronic Engineering, Zhejiang University,
	Hangzhou 310007, China, and Zhejiang Provincial Key Lab of Information Processing, Communication and Networking
	(IPCAN), Hangzhou 310007, China, and the International Joint Innovation Center, Zhejiang University, Haining 314400, China
	(e-mail: chongwenhuang@zju.edu.cn).}
	\thanks{Q. Guo is with the School of Electrical, Computer and Telecommunications Engineering, University of Wollongong, Wollongong, NSW 2522, Australia  (e-mail: qguo@uow.edu.au).}
	\thanks{C. Yuen is with the Engineering Product Development (EPD) Pillar, Singapore University of Technology and
	Design, Singapore 487372 (e-mail: yuenchau@sutd.edu.sg).}
}
\maketitle

\begin{abstract}
Reconfigurable intelligent surface (RIS) is very promising for wireless networks to achieve high energy efficiency, extended coverage, improved capacity, massive connectivity, etc. To unleash the full potentials of RIS-aided communications, acquiring accurate channel state information is crucial, which however is very challenging. %Tremendous efforts have been devoted to the issue of channel estimation of RIS-aided communications. 
For RIS-aided multiple-input and multiple-output (MIMO) communications, the existing channel estimation methods have computational complexity growing rapidly with the number of RIS units $N$ (e.g., in the order of $N^2$ or $N^3$) and/or have special requirements on the matrices involved (e.g., the matrices need to be sparse for algorithm convergence to achieve satisfactory performance), which hinder their applications. 
%normally have a scalability issue as their complexities grow rapidly with the number of RIS units $N$, e.g., in the order of $N^2$ or $N^3$. This makes them inapplicable to RIS with a large or even a moderate number of RIS units. In addition, they normally need a high training overhead to achieve satisfactory estimation performance, which also leads to a high latency, especially in the case of time-varying channels. 
In this work, instead of using the conventional signal model in the literature, we derive a new signal model obtained through proper vectorization and reduction operations. Then, leveraging the unitary approximate message passing (UAMP), we develop a more efficient channel estimator that has complexity linear with $N$ and does not have special requirements on the relevant matrices, thanks to the robustness of UAMP. These facilitate the applications of the proposed algorithm to a general RIS-aided MIMO system with a larger $N$. Moreover, extensive numerical results show that the proposed estimator delivers much better performance and/or requires significantly less number of training symbols, thereby leading to notable reductions in both training overhead and latency. %In addition, the proposed estimator exhibits better robustness against the RIS phase matrix used for training. Extensive simulation results demonstrate the superiority of the proposed channel estimator.
\end{abstract}

\begin{IEEEkeywords}
Reconfigurable intelligent surface (RIS), channel estimation, approximate message passing (AMP). 
\end{IEEEkeywords}

\section{Introduction}
\IEEEPARstart {R}{econfigurable} 
intelligent surface (RIS) is a reconfigurable planar array with a massive number of passive reflecting elements, which is capable of altering the wireless propagation environment to achieve desired channel responses.  % \cite{2020Overhead, 2019Large, 2019LISA, IRS_gaobeiyin_TWC19, 2020BasarReconfigurable, 2020Holographic}. 
RIS has been recognized as a promising technology in future wireless communications. With the aid of RIS, wireless networks are able to achieve high energy efficiency, improve the system capacity and radio coverage, enhance massive connectivity, etc \cite{  2019Reconfigurable, 2019Artificial,  2019Large,  2019LISA, 2020Towards,  IRS_gaobeiyin_TWC19,  2019GLOBECOMGuo,  2020Secrecy,  2020TVTWang, 2020BasarReconfigurable, 2020Holographic, 2020Overhead, 2020YangCoverage,  2021Joint, 2021TWCResource, 2021JSACHuang}. 
%2014Overview, 2018CommunicationsAkyildiz,

The research on RIS has attracted tremendous attention and many works have been conducted to explore the potentials of RIS-aided communications. Under the assumption of perfect channel state information (CSI), energy-efficient designs were studied in \cite{2019Reconfigurable} with the assist of RIS for wireless communications. The works in \cite{2019Artificial} and \cite{2020Secrecy} studied RIS-aided secrecy communications. 
The work in \cite{2021TWCResource} investigated the use of RIS to enhance the energy harvesting and information transmission capabilities of the internet of things systems. Energy efficient transmission with distributed RISs was studied in \cite{IRS_gaobeiyin_TWC19}. The works in \cite{2021Joint} and \cite{2020JSACReconfigurable}  studied RIS-aided non-othogonal multiple access. RIS-aided communications were extended to millimeter-wave and Terahertz communications in \cite{2018CommunicationsAkyildiz, 2021JSACHuang, 2020TVTWang}. The works in  \cite{2019GLOBECOMGuo} and \cite{2020TWCMulticell} maximize the weighted sum rate for both single and multi-cells networks. 
%The work in \cite{2021Joint} investigated to  minimize the total transmit power by jointly optimizing the active beamforming and RIS reflection coefficient. %In order to maximize the achievable secrecy rate, the work in \cite{2019Artificial} designed a method for jointly optimizing transmit beamforming with artificial noise and RIS reflect beamforming. 
%The work in \cite{2020Secrecy} studied the secrecy performance of a RIS-aided wireless communication system in the presence of an eavesdropping user. 
%The work in \cite{2021TWCResource} studied the use of RIS to enhance the energy harvesting and information transmission capabilities of the internet of things systems.
 %The work in \cite{2020JSACReconfigurable} designed a system for serving paired power-domain non-orthogonal multiple access users by designing the passive beamforming weights at RIS. 
% The works in \cite{2020TWCMulticell} and \cite{2019GLOBECOMGuo} maximize the weighted sum rate through  for both single and multi cells networks.}  
To unleash the potentials of RIS aided communications, efficient accurate CSI acquisition is crucial \cite{2020Towards, 2019MishraChannel,2020ZhengIntelligent, 2019AnOptimal,2020NadeemIntelligent,2021TahaEnabling, 2020Cascaded, he2021semiblind, 2020Matrix, 2020WangChannel, 2021Empowered, WeiChau}.

In this work, we focus on channel estimation of RIS-aided multiple-input multiple-output (MIMO) communications, which is challenging especially when the number of RIS unit $N$ is large.
In \cite{2020Cascaded}, a two-stage algorithm that includes a sparse matrix factorization stage and a matrix completion stage was proposed for the estimation of cascaded (transmitter-RIS and RIS-receiver) channel, {where the RIS phase matrix needs to be sparse and the channel between RIS and BS is required to be a low-rank matrix.}  
%\rev{but the feasibility of proposed scheme has strict structural requirements on channel matrices.} 
The work in \cite{he2021semiblind} proposed a trilinear semi-blind cascaded channel estimation problem, in which the receiver estimates the channel coefficients and the transmitted signals jointly, and a message passing algorithm was developed. The algorithm has relatively low complexity, but it also requires that the relevant matrices are sparse for the convergence of the algorithm, thereby achieving good performance. A message-passing algorithm was proposed in \cite{2020Matrix} to estimate the cascaded channels by exploiting the information on the slow-varying channel components and the hidden channel sparsity, and its complexity growing with $N^2$. The sparseness of the matrices may be achieved by using a sparse RIS phase matrix, which however is not optimal in terms of channel estimation and transmission, or by restricting the applications to special scenarios with sparse channel matrices. 
The work in \cite{2020WangChannel} proposed a three-phase pilot-based channel estimation framework for RIS-assisted uplink multiuser communications, and the complexity of the method is cubic in the number of RIS units $N$. %Nevertheless, the running time required by the algorithm can be long in practice if the RIS is equipped with a large number of elements.}
{In \cite{2021Empowered},} leveraging the parallel factor (PARAFAC) decomposition to represent high dimensional tensors that involve unknown channels in different unfolded forms, channel estimation methods based on alternating least squares (ALS) and vector approximate message passing (VAMP) were proposed to estimate the channels from BS to RIS and from RIS to users alternatively. Although there are no requirements on the sparseness of the involved matrices, its complexity grows with $N^3$, which is a concern for a large or even a moderate $N$. In summary, the existing methods have special requirements on the involved matrices and/or have the scalability issue as their computational complexity rapidly grows with $N$, which hinder their applications.

To overcome the problems of the existing methods, leveraging the unitary approximate message passing (UAMP) algorithm \cite{2015GuoApproximate, 2020YuanApproximate, UAMPSBL}, we design a new RIS channel estimator in this work. As a variant of the AMP algorithm \cite{2010DonohoMessage}, UAMP achieves remarkably improved robustness by using a unitary transformation, which enables it to deal with a linear reverse problem with a general (or tough) system transfer matrix while with low complexity \cite{2020YuanApproximate}, \cite{UAMPSBL}.  
%is used in UAMP, which that suffers from has received tremendous attention due to its low complexity to solve a linear inverse problem. However, it may easily diverge for a general system transfer matrix. With a unitary transformation, UAMP \cite{2015GuoApproximate} exhibits remarkably better robustness than AMP \cite{2020YuanApproximate}, \cite{UAMPSBL}. 
In this work, instead of using the conventional signal model for RIS channel estimation in the literature, we derive a new signal model through proper vectorization and reduction operations, which reformulates the channel estimation to a structured signal recovery problem. Then a factor graph representation is developed, and a message passing algorithm is derived, where UAMP plays a crucial role. Thanks to the low complexity and robustness of UAMP, the proposed algorithm is very efficient, which has complexity linear with the number of RIS units $N$, and does not have special requirements on the relevant matrices. These enable the applications of the proposed algorithm to a general RIS aided MIMO system with a larger $N$. The Cramér-Rao lower bound (CRLB) for the channel estimation considered is derived to serve as a performance benchmark. Extensive numerical results show that, with much lower complexity, the proposed channel estimator significantly outperforms existing ones in terms of channel estimation performance and/or the training overhead.

The remainder of the paper is organized as follows. In Section \ref{sec:model}, we introduce the system model and problem formulation of RIS channel estimation. In Section III, the problem is reformulated, a new signal model is obtained and the problem is represented as a factor graph. Then the UAMP based message passing algorithm is developed in Section \ref{sec:Bayesian}, and the CRLB for channel estimation is derived in Section \ref{sec:CRLB}. Numerical results are provided in Section \ref{sec:Simulation}, followed by conclusions in Section \ref{sec:Conclusion}.

\textit{Notations}-Boldface lower-case and upper-case letters denote vectors and matrices, respectively. Superscripts $\boldsymbol{A}^{H}$ and $\boldsymbol{A}^{T}$ represent conjugate transpose and transpose, respectively, and $\boldsymbol{A}^{*}$ represents the conjugate of $\boldsymbol{A}$. A Gaussian distribution of $x$ with mean $\hat{x}$ and variance $\nu_{x}$ is denoted by $\mathcal{N}\left(x ; \hat{x}, \nu_{x}\right).$ Notations $\otimes$ and $\odot$ represent the Kronecker and  Khatri-Rao products, respectively. The relation $f(x)=c g(x)$ for some positive constant $c$ is written as $f(x) \propto g(x)$. We use $\boldsymbol{a} \cdot \boldsymbol{b}$ and $\boldsymbol{a} \cdot / \boldsymbol{b}$ to represent the element-wise product and division between vectors $\boldsymbol{a}$ and $\boldsymbol{b}$, respectively. The notation $\boldsymbol{a}^{.-1}$ denotes the element-wise inverse operation to vector $\boldsymbol{a}$. We use $|\boldsymbol{A}|^{2}$ to denote element-wise magnitude squared operation for $\boldsymbol{A}$, $\|\boldsymbol{a}\|$  to denote the $l_{2}$ norm of $\boldsymbol{a}$, and $\|\boldsymbol{A}\|_{F}$ to denote the Frobenius norm of $\boldsymbol{A}$. The notation $<\boldsymbol{a}>$ denotes the average operation for $\boldsymbol{a}$, i.e., the sum of the elements of $\boldsymbol{a}$ divided by its length. %The notation $\int_{\boldsymbol{c} \vee c_{n}} f_{\boldsymbol{c}}(\boldsymbol{c})$ represents integral over all elements in $\boldsymbol{c}$ except $c_{n}$. 
We use $\mathbf{1}$ and $\mathbf{0}$ to denote an all-one vector and an all-zero vector with a proper length, respectively. %Sometimes, we use a subscript $n$ for $\mathbf{1}$ and $\mathbf{0}$, i.e., $\mathbf{1}_{n}$ and $\mathbf{0}_{n}$ to indicate its length $n$. 
The notation $\textbf{Diag}(\boldsymbol{a})$ represents a diagonal matrix with $\boldsymbol{a}$ as its diagonal and $\boldsymbol{I}_{n}$ donates a $n \times n$ identity matrix. 

\section{System Model and Problem Formulation} \label{sec:model} 
%\subsection{System Model}
\begin{figure}[htbp]
	\centering
	\includegraphics[width=0.9\columnwidth]{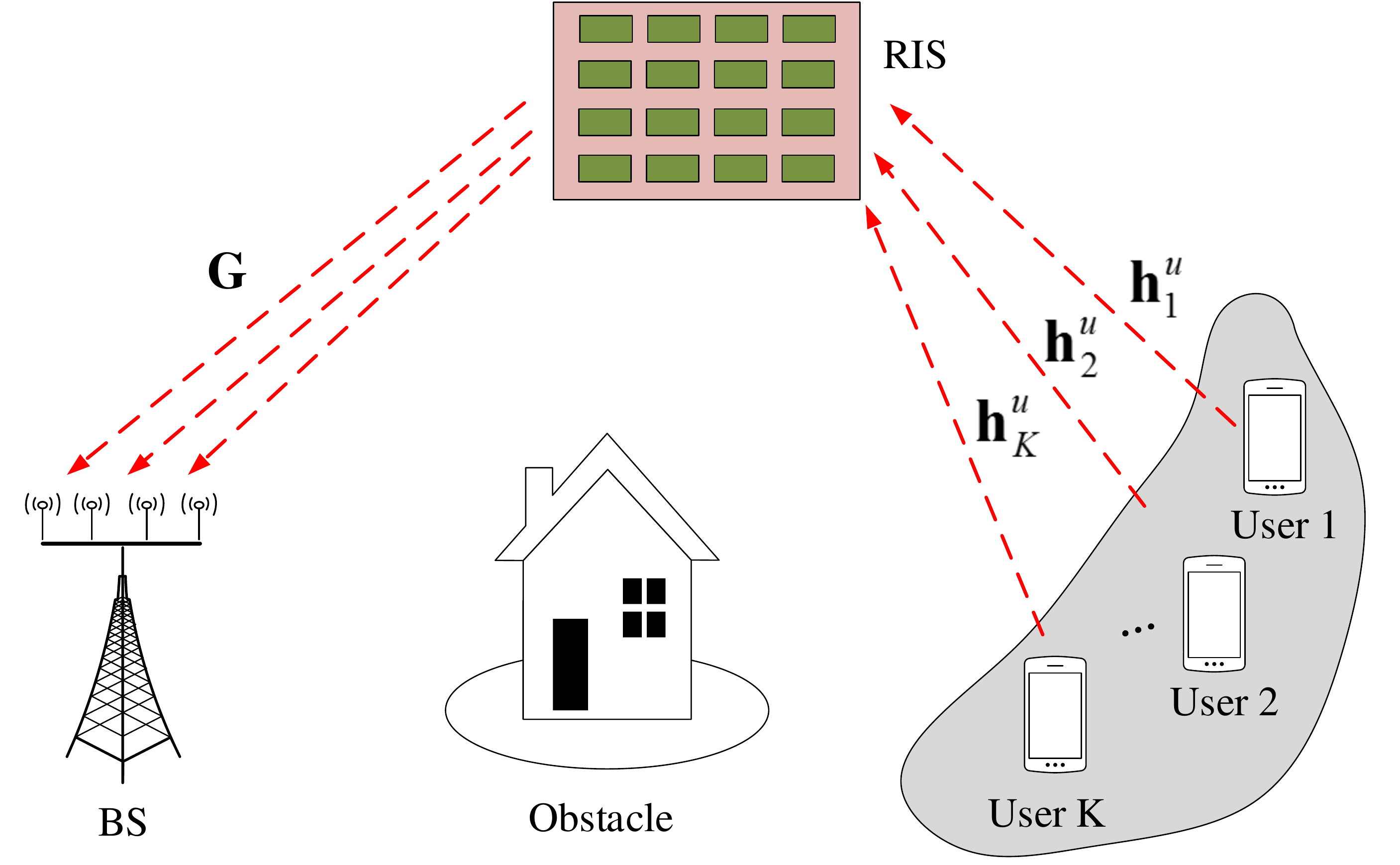}
	\caption{{Illustration of RIS-aided MIMO uplink transmission.}}
	\label{fig:system model}
\end{figure}
We consider RIS-aided MIMO uplink transmission as shown in Fig. \ref{fig:system model}, where the BS equipped with $M$ antennas receives signals from $K$ users. %each equipped with a single antenna. %with the aid of RIS as shown in Fig.\ref{fig:system model}. 
A RIS with $N$ passive reflecting elements is equipped between the BS and users, each user equipped with a single antenna (the extension to multiple antennas is straightforward).
The RIS is attached to the facade of a building in the vicinity of users. Due to the highly attenuation caused by unfavorable propagation environments such as tall buildings, the direct propagation path between the BS and users is neglected \cite{2021Empowered}.
 
{We denote the channel matrix between the BS and the RIS by $\boldsymbol{G} \in \mathbb{C}^{M \times N}$, and use $\boldsymbol{H}\triangleq[\boldsymbol{h}_{1}^{u},\ldots,\boldsymbol{h}_{K}^{u}] \in \mathbb{C}^{N \times K}$ to represent the channel matrix between the RIS and the $K$ users, where $\boldsymbol{h}_{k}^{u}$ denote the channel vector from the $k$-th user to the RIS}. With  the $l$-th ($l =1,2, \ldots,L$) RIS phase configuration, the $K$ consecutive received signals $\boldsymbol{Y}_{l} \in \mathbb{C}^{M \times K}$ is given by %\cite{chen2019channel}
\begin{eqnarray}
 \boldsymbol{Y}_{l} = \boldsymbol{G} \textbf{Diag}(\boldsymbol{\Phi}_{l,:}) \boldsymbol{H} \boldsymbol{X}+\boldsymbol{W}_{l}, \label{eq:recvYl}
\end{eqnarray}
where $\boldsymbol{\Phi}_{l,:}$ is the $l$-th row of the RIS phase matrix $\boldsymbol{\Phi} \in \mathbb{C}^{L \times N}$ (part of the RIS phase configurations) and $\boldsymbol{W}_{l}$ models the zero mean complex additive white Gaussian noise (AWGN) with precision $\beta$ (i.e., variance $\beta^{-1}$), and {$\boldsymbol{X}\in \mathbb{C}^{K \times K}$ denotes the transmitted orthogonal training matrix from the users, i.e., {$\boldsymbol{X}\boldsymbol{X}^H=\boldsymbol{I}_{K}$.} Right-multiplying both sides of  $\eqref{eq:recvYl}$ by $\boldsymbol{X}^H$ leads to}
\begin{eqnarray}
\boldsymbol{\tilde{Y}}_{l}= \boldsymbol{G} \textbf{Diag}(\boldsymbol{\Phi}_{l,:}) \boldsymbol{H}+\boldsymbol{\tilde{W}}_{l}, \label{eq:recvYl_nopilot}
\end{eqnarray}
where $\boldsymbol{\tilde{Y}}_{l} \triangleq \boldsymbol{Y}_{l} \boldsymbol{X}^H  \in \mathbb{C}^{M \times K}$, and $\boldsymbol{\tilde{W}}_{l} \triangleq \boldsymbol{W}_{l} \boldsymbol{X}^H  \in \mathbb{C}^{M \times K}$. 
As $\boldsymbol{X}$ is a unitary matrix, entries of $\boldsymbol{\tilde{W}}_{l}$ are still zero-mean white Gaussian with the same noise precision $\beta$.

{During the training process, $L$ configurations of the RIS are used, leading to $L$ matrices $\{\boldsymbol{\tilde{Y}}_{l}, l=1,..., L\}$, based which we aim to estimate the channel matrices. As the number of RIS unit $N$ can be large, low complexity of channel estimators is crucial for practical implementation. The cubic or quadratic complexity (in terms of $N$) of existing channel estimation algorithms can be a concern in their applications. During the training process, the RIS undergoes $L$ configurations. In order to reduce the training overhead and communication latency, a small $L$ is desirable.}
We address the above challenges by reformulating the RIS channel estimation to a structured signal recovery problem with a new signal model and develop an efficient message passing algorithm. In particular, we do not make any special requirements on the matrices $\boldsymbol{G}$, $\boldsymbol{H}$ or $\boldsymbol{\Phi}$.   

\section{{Problem Reformulation and Factor Graph Representation}}

\subsection{{New Model for RIS Channel Estimation}}
Instead of using model $\eqref{eq:recvYl_nopilot}$ directly, we reformulate it to a new model through vectorization and reduction, which leads to a structured signal recovery problem. Vectorizing $\eqref{eq:recvYl_nopilot}$ leads to
\begin{eqnarray}
\textbf{vec} ( \boldsymbol{\tilde{Y}}_{l}) = ( \boldsymbol{H}^T  \otimes \boldsymbol{G})\textbf{vec}\left(  \textbf{Diag}(\boldsymbol{\Phi}_{l,:})\right)+\textbf{vec}( \boldsymbol{\tilde{W}}_{l}) , \label{eq:vecYl_nopilot}
\end{eqnarray}
{where the vector $\textbf{vec}\left(  \textbf{Diag}(\boldsymbol{\Phi}_{l,:})\right)$ can be represented as
\begin{equation}
\textbf{vec}\left(  \textbf{Diag}(\boldsymbol{\Phi}_{l,:})\right)=[{\phi}_{l,1},\mathbf{0}_{N}^{T},{\phi}_{l,2},\mathbf{0}_{N}^{T},\ldots,\mathbf{0}_{N}^{T},{\phi}_{l,N}]^{T},
\end{equation}
and $\phi_{l,n}$ donates the $n$-th ($n =1,2, \ldots,N$) element of  $\boldsymbol{\Phi}_{l,:}$.} 
{Here we note that, the vector $\textbf{vec}\left(  \textbf{Diag}(\boldsymbol{\Phi}_{l,:})\right)$ is highly sparse, and the non-zeros elements of $\textbf{vec}\left(  \textbf{Diag}(\boldsymbol{\Phi}_{l,:})\right)$ are separated by all-zero vector $\mathbf{0}_{N}$. This can be exploited to significantly reduce the dimension of $\eqref{eq:vecYl_nopilot}$. We drop the zero elements in $\textbf{vec}\left(  \textbf{Diag}(\boldsymbol{\Phi}_{l,:})\right)$ and the corresponding columns in $\boldsymbol{H}^T  \otimes \boldsymbol{G}$, then \eqref{eq:vecYl_nopilot} can be reduced to}
\begin{equation}\label{aftertransform}
	\boldsymbol{\tilde{y}}_{l}=\boldsymbol{\tilde{S}} \boldsymbol{\Phi}^{T}_{l,:}+\boldsymbol{\tilde{w}}_{l},
\end{equation}
{where $\boldsymbol{\tilde{y}}_{l}\triangleq \textbf{vec}(\boldsymbol{\tilde{Y}}_{l})\in \mathbb{C}^{KM \times 1} $, $\boldsymbol{\tilde{S}}\triangleq \boldsymbol{H}^T  \odot \boldsymbol{G}\in \mathbb{C}^{KM \times N} $, $\boldsymbol{\tilde{S}} \boldsymbol{\Phi}^{T}_{l,:}=( \boldsymbol{H}^T  \otimes \boldsymbol{G})\textbf{vec}\left(  \textbf{Diag}(\boldsymbol{\Phi}_{l,:})\right)$, and $\boldsymbol{\tilde{w}}_{l}\triangleq \textbf{vec}( \boldsymbol{\tilde{W}}_{l})\in \mathbb{C}^{KM \times 1}$.
By stacking $\{\boldsymbol{\tilde{y}}_{l}, l=1,...,L\}$ into a matrix, we have}
%\begin{equation}\label{setY}
$\boldsymbol{\tilde{Y}}=\boldsymbol{\tilde{S}}\boldsymbol{\Phi}^{T}+\boldsymbol{\tilde{W}}$,
%\end{equation}
which is rewritten as
\begin{equation}
\boldsymbol{Y}=\boldsymbol{\Phi}\boldsymbol{S}+\boldsymbol{W}. \label{eq:Y_nopilot}
\end{equation}
where $\boldsymbol{Y}=\boldsymbol{\tilde{Y}}^{T}$, $\boldsymbol{W}=\boldsymbol{\tilde{W}}^{T}$and
\begin{equation}\label{eq:S}
	\boldsymbol{S}= \boldsymbol{\tilde{S}}^{T}=(\boldsymbol{H}^T  \odot \boldsymbol{G})^T\in \mathbb{C}^{N \times KM }.
\end{equation}  
{Now we can see from \eqref{eq:Y_nopilot} that, the channel estimation is reformulated as the recovery of the matrix $\boldsymbol{S}$, which admits the structure \eqref{eq:S}. We can also treat $\boldsymbol{S}$ as an intermediate variable, as our aim is to estimate $\boldsymbol{H}$ and $\boldsymbol{G}$.}

\subsection{{Probabilistic and Factor Graph Representation}}\label{sec:Bayesian} 
We consider recovering the signal using the message passing techniques, in particular leveraging UAMP to achieve low complexity while with high robustness. We first represent the problem in a probabilistic form. It is noted that UAMP works with a unitary transform of the linear observation model \eqref{eq:Y_nopilot}, which is crucial to achieving high robustness. So, to facilitate the use of UAMP later, we first carry out a unitary transformation to \eqref{eq:Y_nopilot} based on the singular value decomposition (SVD) $\boldsymbol{\Phi}=\boldsymbol{U \Lambda V}$, leading to 
\begin{equation}\label{svd}
\boldsymbol{R}=\boldsymbol{\Psi}\boldsymbol{S}+\boldsymbol{\overline{W}},
\end{equation}
where {$\boldsymbol{R}=\boldsymbol{U}^{H} \boldsymbol{Y}$, $\boldsymbol{\Psi}=\boldsymbol{U}^{H} \boldsymbol{\Phi}=\boldsymbol{\Lambda V}$ and $\boldsymbol{\overline{W}}=\boldsymbol{U}^{H} \boldsymbol{W}$. }Since $\boldsymbol{U}$ is unitary, entries in $\boldsymbol{\overline{W}}$ are still AWGN with precision $\beta$.

{Note that $\boldsymbol{\tilde{S}}= [\boldsymbol{\tilde{s}}_{1},\ldots,\boldsymbol{\tilde{s}}_{N}]$, $\boldsymbol{H}^{T} = [\boldsymbol{h}_{1},\ldots,\boldsymbol{h}_{N}]$ and $\boldsymbol{G}=[\boldsymbol{g}_{1},\ldots,\boldsymbol{g}_{N}]$. Then, according to $\boldsymbol{\tilde{S}}=\boldsymbol{H}^T \odot \boldsymbol{G}$, we have}
\begin{equation}\label{KR}
	\boldsymbol{\tilde{s}}_{n}=\boldsymbol{h}_{n} \otimes \boldsymbol{g}_{n},
\end{equation}
{where $\boldsymbol{h}_{n}=[{h}_{1,n},\ldots,{h}_{K,n}]^T$, $\boldsymbol{g}_{n}=[{g}_{1,n},\ldots,{g}_{M,n}]^T$ and {$\boldsymbol{\tilde{s}}_{n}=[{\tilde{s}}_{1,1,n},\ldots,{\tilde{s}}_{m,k,n},\ldots,{\tilde{s}}_{M,K,n}]^T$} with ${\tilde{s}}_{m,k,n}={h}_{k,n}{g}_{m,n}$.
Let $J=KM$, and note that $\boldsymbol{R}=[\boldsymbol{r}_{1}, \ldots, \boldsymbol{r}_{J}]$, $\boldsymbol{S}=[\boldsymbol{s}_{1},\ldots,\boldsymbol{s}_{J}]$, and {$\boldsymbol{\overline{W}}=[\boldsymbol{w}_{1},\ldots,\boldsymbol{w}_{J}]$.} Define an auxiliary variable $\boldsymbol{Z}\triangleq[\boldsymbol{z}_{1},\ldots,\boldsymbol{z}_{J}]$ with $\boldsymbol{z}_{j}=\boldsymbol{\Psi}\boldsymbol{s}_{j}$.
Then, the joint distribution of $\boldsymbol{H},\boldsymbol{G},\boldsymbol{S},\boldsymbol{\tilde{S}},\boldsymbol{Z}$ and $\beta$ given $\boldsymbol{R}$ can be factorized as}
\begin{align}
	p&(\boldsymbol{H}, \boldsymbol{G}, \boldsymbol{S}, \boldsymbol{\tilde{S}}, \boldsymbol{Z}, \beta |\boldsymbol{R}) \nonumber \\ 
	\propto &  ~p(\boldsymbol{R}|\boldsymbol{Z},\beta)
	p(\boldsymbol{\tilde{S}}|\boldsymbol{S})
	p(\boldsymbol{Z}|\boldsymbol{S})
	p(\boldsymbol{\tilde{S}}|\boldsymbol{H},\boldsymbol{G})
	p(\boldsymbol{H})p(\boldsymbol{G})
	p(\beta) \nonumber\\
	=&~ p(\boldsymbol{\tilde{S}}|\boldsymbol{S})p(\beta)	
	{\prod}_{n}   
	p(\boldsymbol{\tilde{s}}_{n}|\boldsymbol{h}_{n},\boldsymbol{g}_{n})p(\boldsymbol{h}_{n})p(\boldsymbol{g}_{n})
	{\prod}_{j}
	p(\boldsymbol{r}_{j}|\boldsymbol{z}_{j},\beta)\nonumber \\& \qquad\qquad \times
	p(\boldsymbol{z}_{j}|\boldsymbol{s}_{j})\nonumber \\
	\triangleq &~
	f_{\boldsymbol{S}}(\boldsymbol{\tilde{S}},\boldsymbol{S})
    f_{\beta}(\beta)
	{\prod}_{n}
	f_{\boldsymbol{\tilde{s}}_{n}}(\boldsymbol{\tilde{s}}_{n},\boldsymbol{h}_{n},\boldsymbol{g}_{n})
	f_{\boldsymbol{h}_{n}}(\boldsymbol{h}_{n})
	f_{\boldsymbol{g}_{n}}(\boldsymbol{g}_{n})\nonumber \\ & \qquad\qquad \times
	{\prod}_{j}
	f_{\boldsymbol{r}_{j}}(\boldsymbol{z}_{j},\beta)
	f_{\boldsymbol{z}_{j}}(\boldsymbol{z}_{j},\boldsymbol{s}_j).\label{eq:Factor}
\end{align}
where the involved distributions are listed in Table \ref{tab:factor}. To facilitate the factor graph representation of the factorization in \eqref{eq:Factor}, local functions (factors) are defined, and the correspondence between the distributions and local functions are also shown in Table \ref{tab:factor}. It is noted that  $p(\boldsymbol{H})$ and  $p(\boldsymbol{G})$ represent the priors for the channel matrices $\boldsymbol{H}$ and  $\boldsymbol{G}$, respectively. When no priors are available for the channel matrices, the priors can be set to be a non-informative one, e.g., $\rho_h=\rho_g= +\infty$ in Table \ref{tab:factor}. The factor graph representation is depicted in Fig. \ref{fig:FactorGraph}.
\begin{table}[htb]
	\color{black} 
	\centering
	\renewcommand\arraystretch{1.2}
	\caption{Factors and distributions in (\ref{eq:Factor}).}\label{tab:factor}
	\begin{tabular}{>{\centering}p{30pt}>{\centering}p{60pt} >{\centering \arraybackslash }p{120pt}}
		\hline
		Factor & Distribution & Function  \\
		\hline
		$ f_{\beta}$  & $p(\beta)$ & $\propto \beta^{-1}$ \\	
		$f_{\boldsymbol{h}_{n}}$  & $p(\boldsymbol{h}_{n})$ & $\mathcal{CN}(\boldsymbol{h}_{n}; \boldsymbol{0}_{K}, \rho_h \boldsymbol{I}_{K})$ \\
		$f_{\boldsymbol{g}_{n}}$ & $p(\boldsymbol{g}_{n})$ & $\mathcal{CN}(\boldsymbol{g}_{n}; \boldsymbol{0}_{M}, \rho_g\boldsymbol{I}_{M})$ \\
		$f_{\boldsymbol{r}_{j}}$ & $p(\boldsymbol{r}_{j}|\boldsymbol{z}_{j},\beta)$ & $\mathcal{N}(\boldsymbol{r}_{j};\boldsymbol{z}_{j}, \beta^{-1}\boldsymbol{I}_{L})$ \\
		$f_{\boldsymbol{z}_{j}}$ & $p(\boldsymbol{z}_j|\boldsymbol{s}_j)$ & $\delta(\boldsymbol{z}_{j}-\boldsymbol{\Psi}\boldsymbol{s}_{j})$ \\
		$f_{\boldsymbol{\tilde{s}}_{n}}$ &	$p(\boldsymbol{\tilde{s}}_{n}|\boldsymbol{h}_{n},\boldsymbol{g}_{n})$ & $\delta(\boldsymbol{\tilde{s}}_{n}-\boldsymbol{h}_{n} \otimes \boldsymbol{g}_{n})$ \\		
		$f_{\boldsymbol{S}}$ & $p(\boldsymbol{\tilde{S}}|\boldsymbol{S})$ & $\delta(\boldsymbol{\tilde{S}}^{T}-\boldsymbol{S})$ \\
		\hline
	\end{tabular}
\end{table}

\tikzstyle{factornode} = [draw, fill=white, circle, inner sep=1pt,minimum size=0.8cm]
\tikzstyle{funnode} = [draw, rectangle,fill=black!100, minimum size = 0.6cm]
\begin{figure}[htbp] 
	\centering
	\begin{tikzpicture} [scale=0.8, transform shape]	
		\node (z1)[factornode] at (0,0) {$\boldsymbol{z}_{1}$};
		
		\node (vdot1)[below = 0.5cm of z1]  {$\boldsymbol{\vdots}$};
		
		\node (zJ)[factornode, below = 0.5cm of vdot1]  {$\boldsymbol{z}_{J}$};
		
		%% one column
		\node (fz1)[right = 0.2cm of z1, funnode]  {};
		\node (fzJ)[right = 0.2cm of zJ, funnode]  {};

		% column text
		\node [above= -0.1cm of fz1]{$f_{\boldsymbol{z}_{1}}$};
		\node [above= -0.1cm of fzJ]{$f_{\boldsymbol{z}_{J}}$};
		
		%% two column
		\node (tildes1)[factornode,  right = 0.2cm of fz1]  {$\boldsymbol{s}_{1}$};
		\node (tildesJ)[factornode, right = 0.2cm of fzJ]  {$\boldsymbol{s}_{J}$};
		
		%% one column
		\node (fs1)[funnode, right = 2.2cm of tildes1]  {};
		\node (fsN)[funnode, right =2.2cm of tildesJ]  {};
		
		% column text
		\node [above= -0.1cm of fs1]{$f_{\boldsymbol{\tilde{s}}_{1}}$};
		\node [above= -0.1cm of fsN]{$f_{\boldsymbol{\tilde{s}}_{N}}$};
		\node (s1)[factornode,  left = 0.2cm of fs1]  {$\boldsymbol{\tilde{s}}_{1}$};
		\node (vdot3)[below=0.5cm of s1]  {$\boldsymbol{\vdots}$};
		\node (sN)[factornode, left= 0.2cm of fsN]  {$\boldsymbol{\tilde{s}}_{N}$};	
		%% one column
		\node (g1)[factornode, right = 0.2cm of fs1,yshift=0.6cm, align=center]  {$\boldsymbol{g}_{1}$};
		\node (gN)[factornode, right = 0.2cm of fsN,yshift=-0.8cm, align=center]  {$\boldsymbol{g}_{N}$};
		
		%% one column
		\node (fr1)[funnode, left = 0.2cm of z1]  {};
		\node (frJ)[funnode, left = 0.2cm of zJ]  {};
		\node [above= -0.1cm of fr1]{$f_{\boldsymbol{r}_{1}}$};
		\node [above= -0.1cm of frJ]{$f_{\boldsymbol{r}_{J}}$};
		
		\node (beta)[factornode, left = 0.4cm of fr1,yshift=-1.4cm, align=center]  {$\beta$};
		\node (fbeta)[funnode, left = 0.15cm of beta]  {};
		\node [above= -0.1cm of fbeta]{$f_{\beta}$};

		\node (fS)[funnode, right =0.3cm of tildes1,yshift=-1.4cm, align=center]  {};
		\node [above= -0.1cm of fS]{$f_{\boldsymbol{S}}$};
		
		\node (h1)[factornode,right = 0.2cm of fs1,yshift=-0.6cm, align=center]  {$\boldsymbol{h}_{1}$};
		\node (vdot2)[below = -0.1cm of h1]  {$\boldsymbol{\vdots}$};
		\node (hN)[factornode, right =0.2cm of fs1,yshift=-2.2cm, align=center]  {$\boldsymbol{h}_{N}$};
		\node (fh1)[funnode, right = 0.45cm of h1, align=center]  {};
		\node [above= -0.1cm of fh1]{$f_{\boldsymbol{h}_{1}}$};	
		\node (fg1)[funnode, right =0.45cm of g1]  {};
		\node [above= -0.1cm of fg1]{$f_{\boldsymbol{g}_{1}}$};	
		\node (fhN)[funnode, right = 0.45cm of hN, align=center]  {};
		\node [above= -0.1cm of fhN]{$f_{\boldsymbol{h}_{N}}$};	
		\node (fgN)[funnode, right =0.45cm of gN]  {};
		\node [above= -0.1cm of fgN]{$f_{\boldsymbol{g}_{N}}$};			
		
		\draw (fbeta.east) -- (beta.west);
		\draw(beta.east) -- (fr1.west);
		\draw(fr1.east) -- (z1.west);
		\draw(z1.east) -- (fz1.west);
		\draw(fz1.east) -- (tildes1.west);
		\draw(tildes1.east) -- (fS.west);
		\draw(fS.east) -- (s1.west);			
		\draw(s1.east) -- (fs1.west);	
		\draw (fs1.east) -- (g1.west);
		\draw(g1.east) -- (fg1.west);	
		\draw(fs1.east) -- (h1.west);
		\draw(h1.east) -- (fh1.west);
		
		\draw(beta.east) -- (frJ.west);
		\draw(frJ.east) -- (zJ.west);
		\draw(zJ.east) -- (fzJ.west);
		\draw(fzJ.east) -- (tildesJ.west);
		\draw(tildesJ.east) -- (fS.west);
		\draw(fS.east) -- (sN.west);		
		\draw(sN.east) -- (fsN.west);
		\draw (gN.east) -- (fgN.west);	
		\draw(fsN.east) -- (hN.west);
		\draw(fsN.east) -- (gN.west);
		\draw(hN.east) -- (fhN.west);	
	\end{tikzpicture}
	\caption{ Factor graph representation of \eqref{eq:Factor}.}
	\label{fig:FactorGraph}
\end{figure}
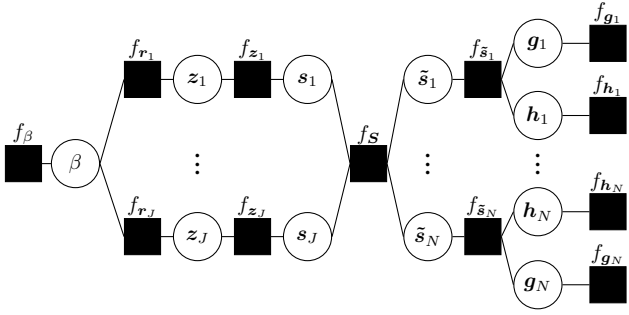

\begin{figure*}[htbp]
	\centering
	\includegraphics[width=1.5\columnwidth]{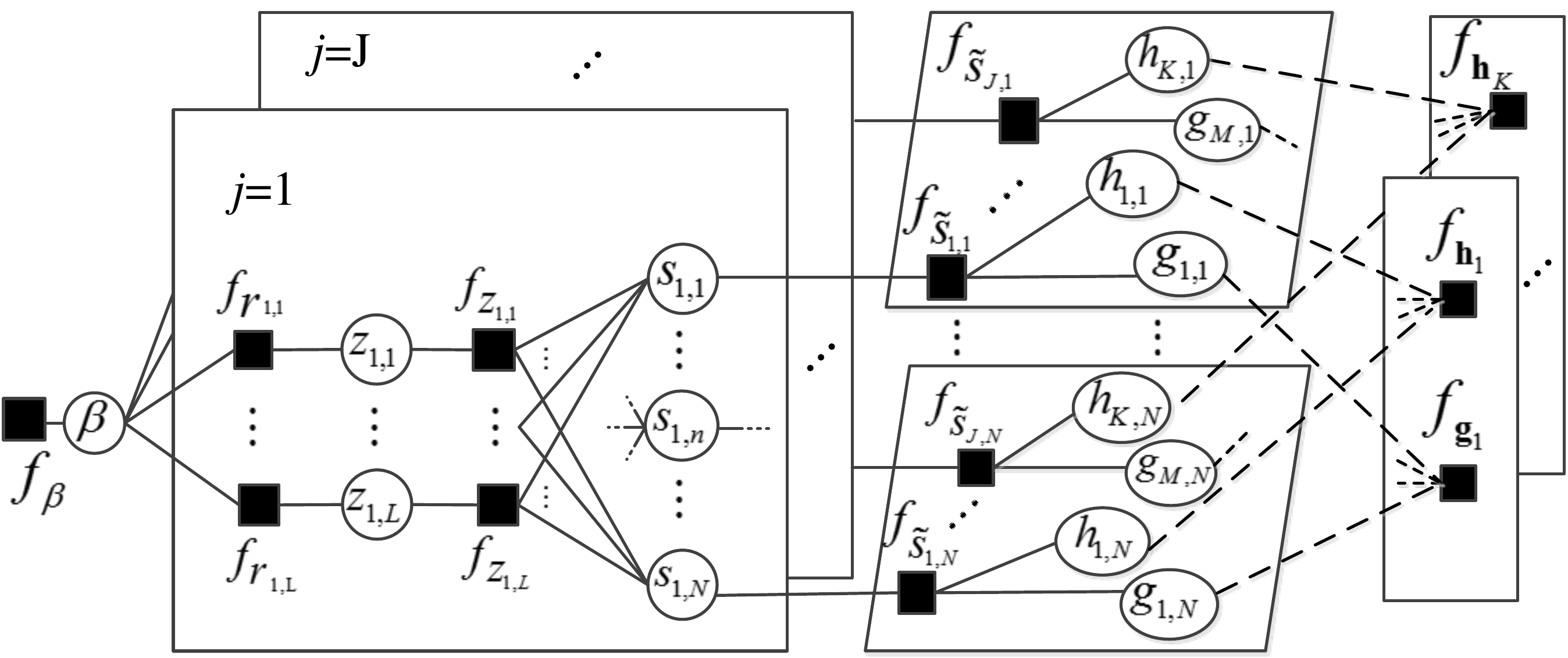}
	\caption{Scalar factor graph representation of \eqref{eq:Factor}.}
	\label{fig:scalarFactorGraph}
\end{figure*}

Our aim is to find the posteriori distributions $p(\boldsymbol{H}|\boldsymbol{R})$ and $p(\boldsymbol{G}|\boldsymbol{R})$ and their estimates in terms of a posteriori means, i.e., $\hat{\boldsymbol{H}}=\mathbb{E}\left\lbrace \boldsymbol{H}|\boldsymbol{R}\right\rbrace $ and $\hat{\boldsymbol{G}}=\mathbb{E}\left\lbrace \boldsymbol{G}|\boldsymbol{R}\right\rbrace $. {We note that it is difficult to find the exact a posteriori distributions, and approximate inference has to be resorted. In next section, we will develop a low complexity message passing algorithm to find their approximations. To facilitate the message passing algorithm design, a scalar factor graph representation is shown in Fig. \ref{fig:scalarFactorGraph}.} 
%The probability function and the corresponding factors are listed in Table \ref{tab:factor}. We can obtain the approximate marginals of the parameters by performing approximate inference. In particular, BiUAMP is derived based on factor graph techniques \cite{2007Variational} and the combined message passing algorithms \cite{2019LuCombined} to achieve efficient approximate inference.

%\begin{figure*}[htbp]
%	\centering
%	\includegraphics[width=1.5\columnwidth]{scalarFactorGraph_v2.pdf}
%	\caption{Scalar factor graph representation of \eqref{eq:Factor}.}
%	\label{fig:scalarFactorGraph}
%\end{figure*}

\section{{UAMP Based Message Passing Algorithm For RIS Channel Estimation}}

{In this section, we develop an efficient message passing algorithm for approximate inference, where UAMP is incorporated to deal with the most computationally intensive part of message passing, which is crucial to achieving low complexity while with high robustness.}
%The factor graph of (\ref{eq:Factor}) can be depicted in Fig. \ref{fig:FactorGraph}.
%It is noted that, all inference methods mentioned above including the structured variational inference (VI), expectation propagation (EP) and UAMP involve an iterative process (but with a different hierarchy), and the multiple iterative processes can be simply combined as a single one. 
{The message passing algorithm carries out an iterative process, where each iteration involves a forward message passing process and backward message passing process in the graph shown in Fig. \ref{fig:FactorGraph} or Fig. \ref{fig:scalarFactorGraph}.}
%In terms of message passing, this is to carry out a forward message passing process and a backward message passing process in Fig. \ref{fig:FactorGraph} as an iteration. For the convenience of understanding, we also give its corresponding scalar factor graph in Fig. \ref{fig:scalarFactorGraph}. 
We use $m_{A\rightarrow B}(x)$ to denote a message passed from node $A$ to node $B$, which is a function of $x$. {For Gaussian messages, the arrows above its mean and variance indicate the direction of the message passing.} In addition, we use $b(x)$ to denote the belief of a variable $x$. Note that, if a forward computation requires backward messages, the relevant messages in the previous iteration is used by default.

\subsection{{Forward Message Passing}}
In the forward direction, according to the rules of variational message passing \cite{2007Variational}, we have
\begin{align}\label{frj_zj}
m_{f_{\boldsymbol{r}_{j}} \rightarrow \boldsymbol{z}_{j}}(\boldsymbol{z}_{j}) & \propto \exp \left\{\int_{\beta} \mathfrak{b}(\beta) \log f_{\boldsymbol{r}_{j}}\right\}\nonumber  \\
& \propto \mathcal{N}\left(\boldsymbol{z}_{j} ; \boldsymbol{r}_{j}, \hat{\beta}^{-1}\right),
\end{align}
where 
\begin{align}\label{beta}
\mathfrak{b}(\beta) & \propto m_{f_{\boldsymbol{r}_{j}} \rightarrow \beta}(\beta) f_{\beta} \nonumber \\
& \propto \beta^{LJ-1} \exp \left\{\sum_{j=1}-\beta\left(\|\boldsymbol{r}_{j}-\hat{\boldsymbol{z}_{j}}\|^{2}+\mathbf{1}^{T} \boldsymbol{\nu}_{\boldsymbol{z}_{j}}\right)\right\},
\end{align}
and 
\begin{equation}\label{betahat}
\hat{\beta}=\int_{\beta} \beta \mathfrak{b}(\beta)=\frac{LJ}{\sum_{j=1}(\|\boldsymbol{r}_{j}-\hat{\boldsymbol{z}_{j}}\|^{2}+\mathbf{1}_{L}^{T} \boldsymbol{\nu}_{\boldsymbol{z}_{j}})}.
\end{equation}
%Here we abuse the use of the notation $\beta$ as we do not distinguish it from the last iteration.  
It is noted that in the above equation, $\hat{\boldsymbol{z}}_{j}$ and the message $m_{ f_{\boldsymbol{r}_{j}} \rightarrow  \beta}( \beta)$  are required to compute $\hat{\beta}$, which are obtained from the last iteration and their computations are delayed to \eqref{frj_beta} and \eqref{zj}. 

%The message $m_{f_{\boldsymbol{r}_{j}} \rightarrow \boldsymbol{z}_{j}}( \boldsymbol{z}_{j})$ is input to \que{the $Part( \uppercase\expandafter{\romannumeral2})$ in Fig. \ref{fig:FactorGraph}.} 
The Gaussian form of the message in \eqref{frj_zj} suggests the following model
\begin{equation}\label{rj}
\boldsymbol{r}_{j}=\boldsymbol{z}_{j}+ \boldsymbol{w}_{j},
\end{equation}
where the noise $\boldsymbol{w}_{j}$ is Gaussian with mean zero and precision $\hat{\beta}$. This allows seamless integration with the forward recursion of UAMP. {Here, assume that the mean and variance of $\boldsymbol{s}_{j}$ are available, i.e., $\hat{\boldsymbol{s}}_{j}$ and $\nu_{\boldsymbol{s}_{j}}$. Specially, we assume  $\boldsymbol{s}_{j}$ have a common variance $\nu_{\boldsymbol{s}_{j}}$, and the corresponding computation will be detailed later. Following UAMP, we define a vector $\boldsymbol{\psi} \in \mathbb{C}^{L \times 1}$ as 
	\begin{equation}\label{psi}
		\boldsymbol{\psi}=|\boldsymbol{\Psi}|^{2}\boldsymbol{1}_{N}.
	\end{equation}
	Then we calculate two vectors $\boldsymbol{\nu_{p_{j}}}$ and $\boldsymbol{p}_{j}$ as
	\begin{equation}\label{vpj}
		\boldsymbol{\nu_{p_{j}}}=\boldsymbol{\psi}\nu_{\boldsymbol{{s}}_{j}},
	\end{equation}
	\begin{equation}\label{pj}
		\boldsymbol{p}_{j}=\boldsymbol{\Psi}\hat{{\boldsymbol{s}}}_{j}-\boldsymbol{\nu_{p_{j}}}\cdot \boldsymbol{\mu}_{j},
	\end{equation}
where $\boldsymbol{\mu}_{j}$ is a vector that is computed in last iteration.} According to UAMP, we update the intermediate vectors $\boldsymbol{\nu}_{\boldsymbol{\mu}_{j}}$ and $\boldsymbol{\mu}_{j}$ by
\begin{align}
\boldsymbol{\nu}_{\boldsymbol{\mu}_{j}}&=\mathbf{1}_{L}. /(\boldsymbol{\nu}_{\boldsymbol{p}_{j}}+\hat{\beta}^{-1} \mathbf{1}_{L}), \label{new_vmuj} \\ 
\boldsymbol{\mu}_{j}&=\boldsymbol{\nu}_{\boldsymbol{\mu}_{j}}\cdot(\boldsymbol{r}_{j}-\boldsymbol{p}_{j}),\label{new_muj}
\end{align}
Then compute vectors $\boldsymbol{\nu}_{\boldsymbol{q}_{j}}$ and $\boldsymbol{q}_{j}$ with
\begin{align}
\boldsymbol{\nu}_{\boldsymbol{q}_{j}} &=\mathbf{1}_{N}./|\boldsymbol{\Psi}^{H}|^{2} \boldsymbol{\nu}_{\boldsymbol{\mu}_{j}},\label{vqj} \\
	\boldsymbol{q}_{j} &=\hat{\boldsymbol{s}}_{j}+\boldsymbol{\nu}_{\boldsymbol{q}_{j}}. \boldsymbol{\Psi}^{H} \boldsymbol{\mu}_{j},\label{qj}
\end{align}
The message $\boldsymbol{q}_{j}$ and $\boldsymbol{\nu}_{\boldsymbol{q}_{j}}$ are the mean and variance of $\boldsymbol{s}_{j}$. According to the belief propagation derivation of (U)AMP, we have
\begin{equation}\label{tlide_sj_fsj}
	m_{\boldsymbol{s}_{j} \rightarrow f_{\boldsymbol{s}}}(\boldsymbol{s}_{j})=\mathcal{N}(\boldsymbol{s}_{j};\boldsymbol{q}_{j},\textbf{Diag}(\boldsymbol{\nu}_{\boldsymbol{q}_{j}})).
\end{equation}
Stack $\boldsymbol{q}_{j}$ and $\boldsymbol{\nu}_{\boldsymbol{q}_{j}}$ into matrices as
\begin{align}\label{Q}
	\boldsymbol{Q} =\left[\boldsymbol{q}_{1}, \ldots, \boldsymbol{q}_{J}\right], 
\end{align}
\begin{align}\label{v_Q}
\boldsymbol{\nu}_{\boldsymbol{Q}} =\left[\boldsymbol{\nu}_{{\boldsymbol{q}}_{1}}, \ldots, \boldsymbol{\nu}_{{\boldsymbol{q}}_{J}}\right],
\end{align}
Due to the deterministic relationship between $\boldsymbol{S}$ and $\boldsymbol{\tilde{S}}$, i.e., $\boldsymbol{\tilde{S}}^{T}=\boldsymbol{S}$, as shown in Table \ref{tab:factor}, we have
%\que{The function node $f_{\boldsymbol{\tilde{S}}}$ is just doing a simple manipulation of the mean and variance of $\boldsymbol{\tilde{S}}$, and we can obtain the mean and variance of $\boldsymbol{S}$ as}
\begin{align}\label{hat_Q}
\boldsymbol{\tilde{Q}}=\boldsymbol{Q}^{T}=\left[\boldsymbol{\tilde{q}}_{1}, \ldots, \boldsymbol{\tilde{q}}_{N}\right], 
\end{align}
\begin{align}\label{V_barQ}
	\boldsymbol{\nu}_{\boldsymbol{\tilde{Q}}}=\boldsymbol{\nu}^{T}_{\boldsymbol{Q}}=[\boldsymbol{\nu}_{\boldsymbol{\tilde{q}}_{1}},\ldots,\boldsymbol{\nu}_{\boldsymbol{\tilde{q}}_{N}}],
\end{align}
and the vectors $\boldsymbol{\tilde{q}}_{n}\in \mathbb{C}^{J\times 1}$ and $\boldsymbol{\nu}_{\boldsymbol{\tilde{q}}_{n}}\in \mathbb{C}^{J\times 1}$ can be divided into $K$ length-$M$ vectors, i.e.,
\begin{equation}\label{qn}
	\boldsymbol{\tilde{q}}_{n}=\left[\boldsymbol{\tilde{q}}^{T}_{1,n}, \ldots, \boldsymbol{\tilde{q}}^{T}_{K,n}\right]^T, 
\end{equation}
\begin{equation}\label{nu_tilde_qn}
 \boldsymbol{\nu}_{\boldsymbol{\tilde{q}}_{n}}=[\boldsymbol{\nu}^{T}_{\boldsymbol{\tilde{q}}_{1,n}}, \ldots, \boldsymbol{\nu}^{T}_{\boldsymbol{\tilde{q}}_{K,n}}]^{T}. 
\end{equation}
Let ${\nu}_{\boldsymbol{\tilde{q}}_{k,n}}=\langle {\boldsymbol{\nu}_{\boldsymbol{\tilde{q}}_{k,n}}}\rangle$, and the message $m_{\boldsymbol{\tilde{s}}_{n} \rightarrow f_{\boldsymbol{\tilde{s}}_{n}}}(\boldsymbol{\tilde{s}}_{n})$ can be expressed as
\begin{equation}\label{sj_fsj}
m_{\boldsymbol{\tilde{s}}_{n} \rightarrow f_{\boldsymbol{\tilde{s}}_{n}}}(\boldsymbol{\tilde{s}}_{n})=\mathcal{N}(\boldsymbol{\tilde{s}}_{n};\boldsymbol{\tilde{q}}_{n},\textbf{Diag}(\boldsymbol{{\nu}}'_{\boldsymbol{\tilde{q}}_{n}})),
\end{equation}
where 
\begin{equation}\label{vqn}
\boldsymbol{{\nu}}'_{\boldsymbol{\tilde{q}}_{n}}=\left[{\nu}_{\boldsymbol{\tilde{q}}_{1,n}}, \ldots, {\nu}_{\boldsymbol{\tilde{q}}_{K,n}}\right]^{T} \otimes \boldsymbol{1}_M.
\end{equation}

%\subsection{\que{Message Computations about $H$ and $G$}} \label{sec:MessageComputations}
Next, we compute the message at $f_{\boldsymbol{\tilde{s}}_{n}}$, $\boldsymbol{h}_{n}$ and $\boldsymbol{g}_{n}$. It is noted that $\boldsymbol{\tilde{s}}_{n}\in \mathbb{C}^{J\times 1}$ also can be divided into $K$ length-$M$ vectors and expressed as
\begin{equation}\label{sn_divided}
	\boldsymbol{\tilde{s}}_{n}=[\boldsymbol{\tilde{s}}_{1,n}^{T},\ldots,\boldsymbol{\tilde{s}}_{K,n}^{T}]^{T},
\end{equation}
where $\boldsymbol{\tilde{s}}_{k,n}\triangleq[{\tilde{s}}_{1,k,n},\ldots, {\tilde{s}}_{M,k,n},]^{T}$. We further factorize the function $f_{\boldsymbol{\tilde{s}}_{n}}(\boldsymbol{\tilde{s}}_{n},\boldsymbol{h}_{n},\boldsymbol{g}_{n})$ as
\begin{equation}\label{f_sn}
f_{\boldsymbol{\tilde{s}}_{n}}(\boldsymbol{\tilde{s}}_{n},\boldsymbol{h}_{n},\boldsymbol{g}_{n})=\prod_{m,k}f_{{\tilde{s}}_{m,k,n}}({h}_{k,n},{g}_{m,n}),
\end{equation}
and the factor $f_{\tilde{s}_{m,k,n}}({h}_{k,n},{g}_{m,n})$ is shown in Fig. \ref{fig:furtherFactorGraph}. % which will be used to derive the forward and backward message computations.

\tikzstyle{factornode} = [draw, fill=white, circle, inner sep=1pt,minimum size=1.2cm]
 \tikzstyle{funnode} = [draw, rectangle,fill=black!100, minimum size = 0.6cm]
 \begin{small}
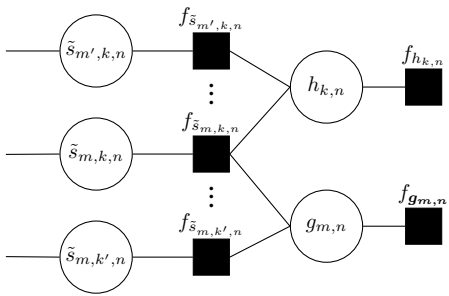
\begin{figure}[htbp] 
	\centering
	\begin{tikzpicture} [scale=0.8, transform shape]	
		\node (upsnkm)[factornode] at (0,0) {$\tilde{s}_{m',k,n}$};
		\node (snkm)[factornode,below = 0.5cm of upsnkm]  {$\tilde{s}_{m,k,n}$};
		\node (downsnkm)[factornode,below = 0.5cm of snkm]  {$\tilde{s}_{m,k',n}$};	
		
		\node (upfsnkm)[funnode, right = 1cm of upsnkm]  {};
		\node (fsnkm)[funnode, right = 1cm of snkm]  {};
		\node (downfsnkm)[funnode, right = 1cm of downsnkm]  {};
		\node [above= -0.1cm of upfsnkm]{$f_{\tilde{s}_{m',k,n}}$};
		\node [above= -0.1cm of fsnkm]{$f_{\tilde{s}_{m,k,n}}$};
		\node [above= -0.1cm of downfsnkm]{$f_{\tilde{s}_{m,k',n}}$};		
		\node (vdot1)[below = -0.1cm of upfsnkm]  {$\boldsymbol{\vdots}$};
		\node (vdot2)[below = -0.1cm of fsnkm]  {$\boldsymbol{\vdots}$};
				\node (hnk)[factornode, right =1cm of upfsnkm,yshift=-0.6cm, align=center]  {$h_{k,n}$};
						\node (gnm)[factornode, below =1cm of hnk,yshift=-0.1cm, align=center]  {$g_{m,n}$};

		\draw[](-1.5,0) -- (upsnkm.west);
		\draw[](upsnkm.east) -- (upfsnkm.west);
 		\draw[](upfsnkm.east) -- (hnk.west);
%		\draw(hnk.east) -- (5,-0.6);
		
		\draw(-1.5,-1.72) -- (snkm.west);
		\draw(snkm.east) -- (fsnkm.west);
 		\draw(fsnkm.east) -- (hnk.west);
 		\draw(fsnkm.east) -- (gnm.west); 	
 			
		\draw[](-1.5,-3.42) -- (downsnkm.west);
		\draw[](downsnkm.east) -- (downfsnkm.west);
 		\draw[](downfsnkm.east) -- (gnm.west);
%		\draw(gnm.east) -- (5,-2.9);	
		
%		\draw($(-1,1)$) rectangle ($(4.8,-4.2)$);
%	 	\draw(0.7,1) -- (0.7,2.5)-- (6.5,2.5)-- (6.5,-2.7)-- (4.8,-2.7);
%	 	% ... 
%	    \node (vdot0)at (0,1.7)  {$\boldsymbol{\cdot}$};
%	 	\node (vdot0)at (0.15,1.8)  {$\boldsymbol{\cdot}$};
%	 	\node (vdot0)at (-0.15,1.6)  {$\boldsymbol{\cdot}$};
%	 	
%	 		 	% ... 
%	 	\node (vdot0)at (5.7,-3.2)  {$\boldsymbol{\cdot}$};
%	 	\node (vdot0)at (5.55,-3.3)  {$\boldsymbol{\cdot}$};
%	 	\node (vdot0)at (5.4,-3.4)  {$\boldsymbol{\cdot}$};	 	
	 	
%	 	\node (smkn1)[factornode]at (1.7,1.7)  {$\tilde{s}_{m',k,1}$};
%	 	\node (fsmkn1)[funnode, right = 1cm of smkn1]  {};
%	 	\node [above= -0.1cm of fsmkn1]{$f_{\tilde{s}_{m',k,1}}$};
%		\node (hkn1)[factornode, right =1cm of fsmkn1,yshift=-0.6cm, align=center]  {$h_{k,1}$};
%		\node (gmn1)[,factornode, below =1cm of hkn1, align=center]  {$g_{m,1}$};
		
\node (hn)[funnode, right = 0.7cm of hnk,yshift=0cm]  {};
\node [above= -0.1cm of hn]{$f_{{h_{k,n}}}$};
	 	
\node (gn)[funnode, right = 0.7cm of gnm]  {};
\node [above= -0.1cm of gn]{$f_{\boldsymbol{g_{m,n}}}$};

%		\draw(0.4,1.7) -- (smkn1.west);
%		\draw(smkn1.east) -- (fsmkn1.west);
%		\draw(fsmkn1.east) -- (hkn1.west);
%		\draw(fsmkn1.east) -- (hkn1.west);
%		\draw[dashed](hkn1.east) -- (hn.west);
		\draw[](hnk.east) -- (hn.west);
%		\draw[dashed](gmn1.east) -- (gn.west);
    	\draw[](gnm.east) -- (gn.west);

	\end{tikzpicture}
	\caption{ Factor graph representation of $f_{{\tilde{s}}_{m,k,n}}$.}
	\label{fig:furtherFactorGraph}
\end{figure}
\end{small}

With the definition $\boldsymbol{\tilde{q}}_{k,n}\triangleq[{\tilde{q}}_{1,k,n},\ldots,{\tilde{q}}_{M,k,n}]^{T}$, \eqref{sj_fsj} implies that 
\begin{align}\label{tilde_s_nmk}
m_{\tilde{s}_{m,k,n} \rightarrow f_{\tilde{s}_{m,k,n}}}\left(\tilde{s}_{m,k,n}\right)=\mathcal{N}\left(\tilde{s}_{m,k,n};\tilde{q}_{m,k,n}, {\nu}_{\boldsymbol{\tilde{q}}_{k,n}}\right), 
\end{align}
and we note that the factor $f_{\tilde{s}_{m,k,n}}=\delta\left(\tilde{s}_{m,k,n}-h_{k,n} g_{m,n}\right)$ as shown in Table \ref{tab:factor}. To compute the message $m_{f_{\tilde{s}_{m,k,n}}\rightarrow g_{m,n}}\left(g_{m,n}\right)$ with belief propagation at factor node $f_{\tilde{s}_{m,k,n}}$, we need to integrate out $s_{m,k,n}$ and $h_{k,n}$. However, due to the multiplication of $g_{m,n}$ and $h_{k,n}$, the message will be intractable even the incoming message $m_{h_{k,n}\rightarrow f_{\tilde{s}_{m,k,n}}}\left(h_{k,n}\right)$ is Gaussian. 
To solve this, we first apply belief propagation and eliminate the variable $\tilde{s}_{m,k,n}$ to get an intermediate function node $\tilde{f}_{\tilde{s}_{m,k,n}}(h_{k,n},g_{m,n})$, i.e.,
\begin{align}\label{tilde_f_s_nmk}
	\tilde{f}_{\tilde{s}_{m,k,n}}(h_{k,n},g_{m,n})&\!=\!\!\int_{\tilde{s}_{m,k,n}}\!\!\!\! m_{\tilde{s}_{m,k,n}\rightarrow f_{\tilde{s}_{m,k,n}}}\!\!\left(\tilde{s}_{m,k,n}\right) \cdot f_{\tilde{s}_{m,k,n}}\nonumber \\
	&=\mathcal{N}\left({\tilde{s}}_{m,k,n}; \tilde{q}_{m,k,n}, {\nu}_{\boldsymbol{\tilde{q}}_{k,n}} \right).
\end{align}
This turns the function node $f_{\tilde{s}_{m,k,n}}$ with the hard constraint $\delta\left(\tilde{s}_{m,k,n}-h_{k,n} g_{m,n}\right)$ to a soft function node, enabling the use of variational inference to handle $h_{k,n}$ and $g_{m,n}$. With the intermediate local function $\tilde{f}_{\tilde{s}_{m,k,n}}(h_{k,n},g_{m,n})$, we can compute the outgoing message $m_{f_{\tilde{s}_{m,k,n}}\rightarrow g_{m,n}}\left(g_{m,n}\right)$ as
\begin{align}\label{f_s_nmk_g_nm}
	m_{f_{\tilde{s}_{m,k,n}}\rightarrow g_{m,n}}\left(g_{m,n}\right)& \propto \exp \left\{\int_{h_{k,n}} \mathfrak{b}\left(h_{k,n}\right) \log \tilde{f}_{\tilde{s}_{m,k,n}}\right\}\nonumber  \\
	&\propto\mathcal{N}(g_{m,n};\rarrow{g}_{m,k,n}, \rarrow{\nu}_{g_{m,k,n}}),
\end{align}
where
\begin{align}\label{right_v_g_nmk}
	\rarrow{\nu}_{g_{m,k,n}} =\frac{{\nu}_{\boldsymbol{\tilde{q}}_{k,n}}}{\left|\hat{h}_{k,n}\right|^{2}+\nu_{h_{k,n}}},
\end{align}
\begin{align}\label{right_g_nmk}
	\rarrow{g}_{m,k,n} =\frac{ \tilde{q}_{m,k,n} \hat{h}_{k,n}^{*}}{\left|\hat{h}_{k,n}\right|^{2}+\nu_{h_{k,n}}},
\end{align}
with $ \hat{h}_{k,n}$ and $\nu_{h_{k,n}}$ being the approximate a posteriori mean and variance of ${h}_{k,n}$, which are computed in (\ref{hat_h_nk}) and (\ref{v_h_nk}). With belief propagation and referring to Fig. \ref{fig:furtherFactorGraph}, the message $m_{g_{m,n}\rightarrow f_{{g_{m,n}}}}\left(g_{m,n}\right)$ can be represented as
\begin{align}\label{g_nm_fgn}
m_{g_{m,n} \rightarrow f_{{g_{m,n}}}}\left(g_{m,n}\right)=\mathcal{N}\left(g_{m,n}; \rarrow{g}_{m,n}, \rarrow{\nu}_{g_{m,n}}\right),
\end{align}
with
\begin{align}\label{right_v_g_nm}
		\rarrow{\nu}_{g_{m,n}} =1/\sum_{k=1}^{K}\frac{1}{\rarrow{\nu}_{g_{m,k,n}}},
\end{align}
\begin{align}\label{right_g_nm}
\rarrow{g}_{m,n} =\rarrow{\nu}_{g_{m,n}} \sum_{k=1}^{K} \frac{\rarrow{g}_{m,k,n}}{\rarrow{\nu}_{g_{m,k,n}}}.
\end{align}
So, the marginal of $g_{m,n}$ can be expressed as 
\begin{align}\label{b_g_nm}
\mathfrak{b}\left(g_{m,n}\right)&= m_{g_{m,n} \rightarrow f_{{g}_{m,n}}}\left(g_{m,n}\right) f_{{g_{m,n}}}\nonumber \\ 
&\propto\mathcal{N}\left(g_{m,n} ; \hat{g}_{m,n}, \nu_{g_{m,n}}\right),
\end{align}
with 
\begin{align}\label{v_g_nm}
	\nu_{g_{m,n}} =\frac{\rarrow{\nu}_{g_{m,n}}\rho_g}{\rho_g+\rarrow{\nu}_{g_{m,n}}},
\end{align}
\begin{align}\label{hat_g_nm}
	\hat{g}_{m,n} =\frac{\rarrow{g}_{m,n} \rho_g}{\rho_g+\rarrow{\nu}_{g_{m,n}}}.
\end{align}
Similarly, we can compute the message $m_{f_{\tilde{s}_{m,k,n}}\rightarrow h_{k,n}}$ as
\begin{align}\label{f_s_nmk_h_nk}
	m_{f_{\tilde{s}_{m,k,n}}\rightarrow h_{k,n}}\left(h_{k,n}\right)\propto \mathcal{N}\left(h_{k,n};\rarrow{h}_{m,k,n}, \rarrow{\nu}_{h_{m,k,n}}\right),
\end{align}
where
\begin{align}\label{right_v_h_nmk}
	\rarrow{\nu}_{h_{m,k,n}} =\frac{{\nu}_{\boldsymbol{\tilde{q}}_{k,n}}}{\left|\hat{g}_{m,n}\right|^{2}+\nu_{g_{m,n}}},
\end{align}
\begin{align}\label{right_h_nmk}
	\rarrow{h}_{m,k,n} =\frac{ \tilde{q}_{m,k,n} \hat{g}_{m,n}^{*}}{\left|\hat{g}_{m,n}\right|^{2}+\nu_{g_{m,n}}},
\end{align}
with $ \hat{g}_{m,n}$ and $\nu_{g_{m,n}}$ being the approximate a posteriori mean and variance of ${g}_{m,n}$, which are computed in (\ref{hat_g_nm}) and (\ref{v_g_nm}). With belief propagation, the message $m_{h_{k,n}\rightarrow f_{{h_{k,n}}}}\left(h_{k,n}\right)$ can be represented as
\begin{align}\label{h_nk_fhn}
	m_{h_{k,n} \rightarrow f_{{h_{k,n}}}}\left(h_{k,n}\right)=\mathcal{N}\left(h_{k,n}; \rarrow{h}_{k,n}, \rarrow{\nu}_{h_{k,n}}\right),
\end{align}
with
\begin{align}\label{right_v_h_nk}
	\rarrow{\nu}_{h_{k,n}} =1/\sum_{m=1}^{M}\frac{1}{\rarrow{\nu}_{h_{m,k,n}}},
\end{align}
\begin{align}\label{right_h_nk}
\rarrow{h}_{k,n} =\rarrow{\nu}_{h_{k,n}} \sum_{m=1}^{M} \frac{\rarrow{h}_{m,k,n}}{\rarrow{\nu}_{h_{m,k,n}}}.
\end{align}
The marginal of $h_{k,n}$ can be expressed as
\begin{align}\label{b_h_nk}
	\mathfrak{b}\left(h_{k,n}\right)&= m_{h_{k,n} \rightarrow f_{{h}_{k,n}}}\left(h_{k,n}\right) f_{{h_{k,n}}}\nonumber \\
	&\propto\mathcal{N}\left(h_{k,n} ; \hat{h}_{k,n}, \nu_{h_{k,n}}\right),
\end{align}
with
\begin{align}\label{v_h_nk}
\nu_{h_{k,n}} =\frac{\rarrow{\nu}_{h_{k,n}}\rho_h}{\rho_h+\rarrow{\nu}_{h_{k,n}}}, 
\end{align}
\begin{align}\label{hat_h_nk}
	\hat{h}_{k,n} =\frac{\rarrow{h}_{k,n}\rho_h}{\rho_h+\rarrow{\nu}_{h_{k,n}}}.
\end{align}
This is the end of forward message passing.

\subsection{Backward message passing}
Next, we elaborate the backward message passing. The backward message from $h_{k,n}$ to $f_{\tilde{s}_{m,k,n}}$ can be expressed as
\begin{align}\label{h_nk_f_s_nmk}
m_{h_{k,n}\rightarrow f_{\tilde{s}_{m,k,n}}}\left(h_{k,n}\right) =\frac{\mathfrak{b}\left(h_{k,n}\right)}{m_{f_{\tilde{s}_{m,k,n} \rightarrow h_{k,n}}}\left(h_{k,n}\right)},
\end{align}
They are represented collectively as $m_{\boldsymbol{h}_{n}\rightarrow f_{\boldsymbol{\tilde{s}}_{n}}}\left(\boldsymbol{h}_{n}\right) $, which is Gaussian with mean $\larrow{\boldsymbol{h}}_n$ and variance $\textbf{Diag}(\larrow{\boldsymbol{\nu}}_{\boldsymbol{h}_n})$. With the factor graph shown in Fig. \ref{fig:furtherFactorGraph}, the mean and variance can be computed as
\begin{align}\label{left_v_hn}
\larrow{\boldsymbol{\nu}}_{\boldsymbol{h}_n}=\left(\left(\mathbf{1}_{K} . / \boldsymbol{\nu}_{\boldsymbol{h}_n}\right) \otimes \mathbf{1}_{M}-\mathbf{1}_{J}. / \rarrow{\boldsymbol{\nu}}_{\boldsymbol{h}_n}\right)^{.-1},
\end{align}
\begin{align}\label{left_hn}
\larrow{\boldsymbol{h}}_n=\larrow{\boldsymbol{\nu}}_{\boldsymbol{h}_n} \cdot \left((\hat{\boldsymbol{h}}_n ./ \boldsymbol{\nu}_{\boldsymbol{h}_n}) \otimes \mathbf{1}_{M}-\rarrow{\boldsymbol{h}}_n ./ \rarrow{\boldsymbol{\nu}}_{\boldsymbol{h}_n}\right),
\end{align}
where $ \boldsymbol{\nu}_{\boldsymbol{h}_n}=[\nu_{h_{1,n}},\ldots,\nu_{h_{K,n}}]^{T} $, $\hat{ \boldsymbol{h}}_n=[\hat{h}_{1,n},\ldots,\hat{h}_{K,n}]^{T} $, $\left[ \rarrow{\boldsymbol{\nu}}_{\boldsymbol{h}_n}\right]_{(k-1)M+m}=\rarrow{\nu}_{h_{m,k,n}} $ and $\left[ {\rarrow{\boldsymbol{h}}_n}\right] _{(k-1)M+m}=\rarrow{h}_{m,k,n} $.

Similarly, the message $m_{\boldsymbol{g}_{n}\rightarrow f_{\boldsymbol{s}_{n}}}\left(\boldsymbol{g}_{n}\right) $ is also Gaussian with mean $\larrow{\boldsymbol{g}}_n$ and variance $\textbf{Diag}(\larrow{\boldsymbol{\nu}}_{\boldsymbol{g}_n})$, which can be computed as
\begin{align}\label{left_v_gn}
	\larrow{\boldsymbol{\nu}}_{\boldsymbol{g}_n}=\left(\mathbf{1}_{K}\otimes \left(\mathbf{1}_{M} . / \boldsymbol{\nu}_{\boldsymbol{g}_n}\right)  -\mathbf{1}_{J}. / \rarrow{\boldsymbol{\nu}}_{\boldsymbol{g}_n}\right)^{.-1},
\end{align}
\begin{align}\label{left_gn}
	\larrow{\boldsymbol{g}}_n=\larrow{\boldsymbol{\nu}}_{\boldsymbol{g}_n} \cdot \left( \mathbf{1}_{K} \otimes (\hat{\boldsymbol{g}}_n ./ \boldsymbol{\nu}_{\boldsymbol{g}_n}) -\rarrow{\boldsymbol{g}}_n ./ \rarrow{\boldsymbol{\nu}}_{\boldsymbol{g}_n}\right),
\end{align}
{where $\boldsymbol{\nu}_{\boldsymbol{g}_n}=[{\nu}_{{g}_{1,n}},\ldots,{\nu}_{{g}_{M,n}}]^T$, $\hat{\boldsymbol{g}}_{n}=[\hat{g}_{1,n},\ldots,\hat{g}_{M,n}]^T $,} $\left[ \rarrow{\boldsymbol{\nu}}_{\boldsymbol{g}_n}\right]_{(m-1)K+k}=\rarrow{\nu}_{g_{m,k,n}}$ and $\left[ {\rarrow{\boldsymbol{g}}_n}\right] _{(m-1)K+k}=\rarrow{g}_{m,k,n} $. Then, the backward message $ m_{f_{\boldsymbol{\tilde{s}}_{n}}\rightarrow \boldsymbol{\tilde{s}}_{n}}(\boldsymbol{\tilde{s}}_{n})=\mathcal{N}(\boldsymbol{\tilde{s}}_{n}; \larrow{\boldsymbol{\tilde{s}}}_{n}, \larrow{\boldsymbol{\nu}}_{\boldsymbol{\tilde{s}}_{n}})$ with
{\begin{align}\label{left_sn}
\larrow{\boldsymbol{\tilde{s}}}_{n}=\larrow{\boldsymbol{h}}_{n} \cdot \larrow{\boldsymbol{g}}_{n},
\end{align}
\begin{align}\label{left_v_sn}
\larrow{\boldsymbol{\nu}}_{\boldsymbol{\tilde{s}}_{n}}=|\larrow{\boldsymbol{h}}_{n}|^{2} \cdot \larrow{\boldsymbol{\nu}}_{\boldsymbol{g}_n}+\larrow{\boldsymbol{\nu}}_{\boldsymbol{h}_n} \cdot| \larrow{\boldsymbol{g}}_{n}|^{2}+\larrow{\boldsymbol{\nu}}_{\boldsymbol{h}_n} \cdot \larrow{\boldsymbol{\nu}}_{\boldsymbol{g}_n},
\end{align}where $\larrow{\boldsymbol{\tilde{s}}}_{n}=[\larrow{\boldsymbol{\tilde{s}}}_{1,n}^{T},\ldots,\larrow{\boldsymbol{\tilde{s}}}_{K,n}^{T}]^T$} and $\larrow{\boldsymbol{\nu}}_{\boldsymbol{\tilde{s}}_{n}}=[\larrow{\boldsymbol{\nu}}_{\boldsymbol{\tilde{s}}_{1,n}}^{T},\ldots,\larrow{\boldsymbol{\nu}}_{\boldsymbol{\tilde{s}}_{K,n}}^{T}] ^{T} $.
Following (U)AMP, the backward message is combined with message $m_{\boldsymbol{\tilde{s}}_{n} \rightarrow f_{\boldsymbol{\tilde{s}}_{n}}}\left(\boldsymbol{\tilde{s}}_{n}\right) $, i.e.,
\begin{align}\label{v_snk}
	\boldsymbol{\nu}_{\boldsymbol{\tilde{s}}_{k,n}} =\left(1 / \nu_{\boldsymbol{q}_{k,n}} \mathbf{1}_{M}+\mathbf{1}_{M}./\larrow{\boldsymbol{\nu}}_{\boldsymbol{\tilde{s}}_{k,n}}\right)^{.-1},
\end{align}
\begin{align}\label{hat_snk}
	\hat{\boldsymbol{\tilde{s}}}_{k,n}=\boldsymbol{\nu}_{\boldsymbol{\tilde{s}}_{k,n}} \cdot \left(1 / \nu_{\boldsymbol{\tilde{s}}_{k,n}} \hat{\boldsymbol{q}}_{k,n}+\larrow{\boldsymbol{\tilde{s}}}_{k,n} ./ \larrow{\boldsymbol{\nu}}_{\boldsymbol{\tilde{s}}_{k,n}}\right),
\end{align}
Stack $\boldsymbol{\nu}_{\boldsymbol{\tilde{s}}_{k,n}}$ and $\hat{\boldsymbol{\tilde{s}}}_{k,n}$ into 
\begin{align}\label{nu_s}
	\boldsymbol{\nu}_{\boldsymbol{\tilde{S}}}=[\boldsymbol{\nu}_{\boldsymbol{\tilde{s}}_{1}},\ldots,\boldsymbol{\nu}_{\boldsymbol{\tilde{s}}_{N}}],
\end{align}
\begin{align}\label{hat_s}
	\hat{\boldsymbol{\tilde{S}}}=[\hat{\boldsymbol{\tilde{s}}}_{1},\ldots, \hat{\boldsymbol{\tilde{s}}}_{N}],
\end{align}
where
\begin{align}\label{nu_s_n}
\boldsymbol{\nu}_{\boldsymbol{\tilde{s}}_{n}}=[\boldsymbol{\nu}_{\boldsymbol{\tilde{s}}_{1,n}}^{T},\ldots,\boldsymbol{\nu}_{\boldsymbol{\tilde{s}}_{K,n}}^{T}]^{T},
\end{align}
{\begin{align}\label{hat_s_n}
\hat{\boldsymbol{\tilde{s}}}_{n}=[\hat{\boldsymbol{\tilde{s}}}_{1,n}^{T},\ldots, \hat{\boldsymbol{\tilde{s}}}_{K,n}^{T}]^{T}.
\end{align} }
Further, we can obtain $\boldsymbol{\nu}_{\boldsymbol{S}}=\boldsymbol{\nu}_{\boldsymbol{\tilde{S}}}^{T}$ and $\hat{\boldsymbol{S}}=\hat{\boldsymbol{\tilde{S}}}^{T}$ with
\begin{align}\label{last_tilde_vs}
 \boldsymbol{\nu}_{\boldsymbol{S}}=[ \boldsymbol{\nu}_{{\boldsymbol{s}}_1},\ldots, \boldsymbol{\nu}_{{\boldsymbol{s}}_J}],
\end{align}
\begin{align}\label{last_average_tilde_vs}
{\nu}_{\boldsymbol{s}_{j}}=\langle {\boldsymbol{\nu}_{\boldsymbol{s}_j}}\rangle,
\end{align}
 \begin{align}\label{last_tilde_s}
\hat{\boldsymbol{S}}=[\hat{\boldsymbol{s}}_1,\ldots,\hat{{\boldsymbol{s}}}_J],
 \end{align}
 \begin{align}\label{last_s}
	\hat{{\boldsymbol{s}}}_j=[\hat{s}_{j,1},\ldots,\hat{s}_{j,N}],
\end{align}

According to the belief propagation derivation of (U)AMP, it holds that
\begin{equation}\label{zj_frj}
	m_{\boldsymbol{z}_{j} \rightarrow f_{\boldsymbol{r}_{j}}}(\boldsymbol{z}_{j})=m_{f_{\boldsymbol{z}_{j}}\rightarrow\boldsymbol{z}_{j}}(\boldsymbol{z}_{j})=\mathcal{N}(\boldsymbol{z}_{j};\boldsymbol{p}_{j},\textbf{Diag}(\boldsymbol{\nu_{p_{j}}})),
\end{equation}
{where $\boldsymbol{p}_{j}$ and $\boldsymbol{\nu_{p_{j}}}$ are computed in the forward direction.} It is noted that the factor node $f_{\boldsymbol{r}_{j}}$ connects the variable node $\beta$. According to the rules of the variational message passing, the message $m_{ f_{\boldsymbol{r}_{j}} \rightarrow \beta}( \beta )$ can be expressed as
\begin{equation}\label{frj_beta}
	m_{f_{\boldsymbol{r}_{j}} \rightarrow \beta}(\beta) \propto \exp \left\{\sum_{j=1}^{J}\int_{\boldsymbol{z}_{j}} \mathfrak{b}(\boldsymbol{z}_{j}) \log f_{\boldsymbol{r}_{j}}\right\},
\end{equation}
where $ \mathfrak{b}(\boldsymbol{z}_{j})$ is the approximate marginal of $\boldsymbol{z}_{j}$, which can be expressed as
\begin{align}
	\mathfrak{b}(\boldsymbol{z}_{j}) & \propto m_{f_{\boldsymbol{r}_{j}} \rightarrow \boldsymbol{z}_{j}}(\boldsymbol{z}_{j}) m_{\boldsymbol{z}_{j} \rightarrow f_{\boldsymbol{r}_{j}}}(\boldsymbol{z}_{j}) \nonumber \\
	&=\mathcal{N}\left(\boldsymbol{z}_{j} ; \hat{\boldsymbol{z}}_{j}, \textbf{Diag}\left(\boldsymbol{\nu}_{\boldsymbol{z}_{j}}\right)\right) ,
\end{align}
where
\begin{equation}\label{vzj}
	\boldsymbol{\nu}_{\boldsymbol{z}_{j}} =\mathbf{1}_{L}./\left(\mathbf{1}_{L} \cdot / \boldsymbol{\nu}_{\boldsymbol{p}_{j}}+\hat{\beta} \mathbf{1}_{L}\right),
\end{equation}
\begin{equation}\label{zj}
	\hat{\boldsymbol{z}}_{j} =\boldsymbol{\nu}_{\boldsymbol{z}_{j}} \cdot\left(\boldsymbol{p}_{j} \cdot / \boldsymbol{\nu}_{\boldsymbol{p}_{j}}+\hat{\beta} \boldsymbol{r}_{j}\right),
\end{equation}
with $\hat{\beta}$ being the approximate a posteriori mean of the noise precision that is obtained with \eqref{betahat}. It is noted that in the above derivation, the message $m_{ f_{\boldsymbol{r}_{j}} \rightarrow \boldsymbol{z}_{j}}(  \boldsymbol{z}_{j} )$  is required, which is Gaussian, i.e., $m_{ f_{\boldsymbol{r}_{j}} \rightarrow \boldsymbol{z}_{j}}(  \boldsymbol{z}_{j} )=\mathcal{N}(\boldsymbol{z}_{j};\boldsymbol{r}_{j},\hat{\beta}^{-1})$ , and its derivation is shown in \eqref{frj_zj}. Then, it is not hard to show that the message
\begin{equation}\label{new_frj_beta}
	m_{f_{\boldsymbol{r}_{j}} \rightarrow \beta}(\beta) \propto \beta^{LJ} \exp \left\{\sum^{J}_{j=1}-\beta\left(\|\boldsymbol{r}_{j}-\hat{\boldsymbol{z}_{j}}\|^{2}+\mathbf{1}_{L}^{T} \boldsymbol{\nu}_{\boldsymbol{z}_{j}}\right)\right\}.
\end{equation}
This is the end of backward message passing.

The message passing algorithm is summarized in Algorithm \ref{algorithm} and it can be terminated when it reaches a maximum number of iteration or the difference between the estimates of two consecutive iterations is less than a threshold.

\begin{algorithm}
	\setstretch{1.25}
	\caption{UAMP-Based Channel Estimation for RIS-Aided MIMO System}
	\textbf{Input:}  A feasible $\boldsymbol{\Phi}$, $\epsilon >0 $ and the maximum number of iteration $I_{max}$.
	\\
	\textbf{Initialize:}  $\hat{h}_{k,n}, \nu_{h_{k,n}}=1, \hat{{\boldsymbol{s}}}_{j}=\boldsymbol{0}, {\nu}_{{\boldsymbol{s}}_{j}}=1, \boldsymbol{\mu}_{j}=\boldsymbol{0}$, $\forall k,n,j$, and $\hat{\beta}=1$. 
	\\
	\textbf{Repeat:}
	\begin{algorithmic}[1]
		\STATE update noise precision $\hat{\beta}$ with (\ref{betahat});{\\
		\STATE $\forall j$: update $\boldsymbol{\nu_{p_{j}}}$ and $\boldsymbol{p}_{j}$ with (\ref{vpj}) and (\ref{pj}); }\\
		\STATE $\forall j$: update $\boldsymbol{\nu}_{\boldsymbol{\mu}_{j}}$ and $\boldsymbol{\mu}_{j}$ with (\ref{new_vmuj}) and (\ref{new_muj});\\
		\STATE $\forall j$: update $\boldsymbol{\nu}_{\boldsymbol{q}_{j}}$ and $\boldsymbol{q}_{j}$ with (\ref{vqj}) and (\ref{qj});\\				
		\STATE $\forall n$: update $\boldsymbol{{\nu}}'_{\boldsymbol{\tilde{q}}_{n}}$ and $ \boldsymbol{\tilde{q}}_{n}$ with (\ref{vqn}) and (\ref{qn});\\
		\STATE $\forall m,k,n$: update $\rarrow{\nu}_{g_{m,k,n}}$ and $\rarrow{g}_{m,k,n} $ with (\ref{right_v_g_nmk}) and (\ref{right_g_nmk});\\
		\STATE $\forall m,n$: update $\rarrow{\nu}_{g_{m,n}} $ and $\rarrow{g}_{m,n}$ with (\ref{right_v_g_nm}) and (\ref{right_g_nm});\\	
		\STATE $\forall m,n$: update $	\nu_{g_{m,n}}$ and $\hat{g}_{m,n}$ with (\ref{v_g_nm}) and (\ref{hat_g_nm}); \\		
		\STATE $\forall m,k,n$: update $\rarrow{\nu}_{h_{m,k,n}}$ and $\rarrow{h}_{m,k,n}$ with (\ref{right_v_h_nmk}) and (\ref{right_h_nmk}); \\    		   
		\STATE $\forall k,n$: update $\rarrow{\nu}_{h_{k,n}}$ and $\rarrow{h}_{k,n}$ with (\ref{right_v_h_nk}) and (\ref{right_h_nk});\\
    	\STATE $\forall k,n$: update $\nu_{h_{k,n}}$ and $	\hat{h}_{k,n}$ with (\ref{v_h_nk}) and (\ref{hat_h_nk});\\
		\STATE $\forall n$: update $\larrow{\boldsymbol{\nu}}_{\boldsymbol{h}_n}$ and $\larrow{\boldsymbol{h}}_n$ with (\ref{left_v_hn}) and (\ref{left_hn});\\
		\STATE $\forall n$: update $\larrow{\boldsymbol{\nu}}_{\boldsymbol{g}_n}$ and $	\larrow{\boldsymbol{g}}_n$ with (\ref{left_v_gn}) and (\ref{left_gn});\\	
		\STATE $\forall n$: update $\larrow{\boldsymbol{\nu}}_{\boldsymbol{\tilde{s}}_n}$ and $\larrow{\boldsymbol{\tilde{s}}}_n$ with (\ref{left_v_sn}) and (\ref{left_sn});\\
		\STATE $\forall k,n$: update $\boldsymbol{\nu}_{\boldsymbol{\tilde{s}}_{k,n}}$ and $\hat{\boldsymbol{\tilde{s}}}_{k,n}$ with (\ref{v_snk}) and (\ref{hat_snk});\\
		\STATE $\forall j$: update ${\nu}_{{\boldsymbol{s}}_{j}}$ and ${\hat{\boldsymbol{s}}_j}$ with (\ref{last_average_tilde_vs}) and (\ref{last_s});\\
		\STATE $\forall j$: update $\boldsymbol{\nu}_{\boldsymbol{z}_{j}}$ and $\hat{\boldsymbol{z}}_{j}$ with (\ref{vzj}) and (\ref{zj}); \\
		\STATE Construct $\hat{\boldsymbol{H}}=[\hat{\boldsymbol{h}}_1,\ldots,\hat{\boldsymbol{h}}_N]^T$ with $\hat{\boldsymbol{h}}_n=[\hat{{h}}_{1,n},\ldots,\hat{{h}}_{K,n}]^T$ and $\hat{\boldsymbol{G}}=[\hat{\boldsymbol{g}}_1,\ldots,\hat{\boldsymbol{g}}_N]$ with $\hat{\boldsymbol{g}}_n=[\hat{{g}}_{1,n},\ldots,\hat{{g}}_{M,n}]^T$.
		\end{algorithmic}
\noindent \textbf{Until} $|| \hat{\boldsymbol{H}}-\boldsymbol{H}||_{F}^{2}\left \| \boldsymbol{H}\right \|_{F}^{-2}< \epsilon$ and $||\hat{\boldsymbol{G}}-\boldsymbol{G}||_{F}^{2}\left \| \boldsymbol{G}\right \|_{F}^{-2}< \epsilon$ or the number of iteration is more than $I_{max}$.

\textbf{Output:} $\hat{\beta}$, $\hat{\boldsymbol{H}}$ and $\hat{\boldsymbol{G}}$ that are the estimations of ${\beta}$, ${\boldsymbol{H}}$ and ${\boldsymbol{G}}$, respectively.
	\label{algorithm}
\end{algorithm}

\subsection{Computational Complexity Analysis}
{We analyze the computational complexity of the proposed algorithm and compare it with that of sate-of-the-art algorithms. The UAMP-based message  passing algorithm needs pre-processing,} i.e., performing a single economic SVD for  $\boldsymbol{\Phi}$ and unitary transformation, and the complexity is $\mathcal{O}(NL^{2})$.
{It noted that the {SVD} can be carried out offline and there is no matrix inversion involved in Algorithm \ref{algorithm}. Also note that the formulated problem is a multiple measurement vector one. The complexity of the proposed algorithm is dominated by the computation of $\boldsymbol{p}$ in step 1, which requires $\mathcal{O}(NLKM+LKM)$, and the computations of  $\nu_{\boldsymbol{q}}$ and $\boldsymbol{q}$ in step 6, which require $\mathcal{O}(NLKM+NKM)$ and $\mathcal{O}(NLKM)$, respectively. It can be shown that the overall complexity of the algorithm is $\mathcal{O}(NLKM)+ \mathcal{O}(NKM)$ per iteration.} 
As we consider a general RIS-aided MIMO system without any special requirements on the (channel or RIS phase) matrices, the most relevant algorithms for comparison are the ALS-based algorithm and VAMP-based algorithm in \cite{2021Empowered}. {The complexity of the ALS-based algorithm is $\mathcal{O}(NLKM+L(M+K)N^{2})+ \mathcal{O}(N^{3})$ per iteration, and that of the VAMP-based algorithm is $\mathcal{O}((M+K)N^{3}+(M+K)N^{2})$ per iteration. It is noted that $N$ is the number of RIS units, which can be much larger than $M$, $K$ and $L$. From the analysis, we can see that the complexity of the proposed algorithm, which is linear with $N$, is significantly smaller than that of the ALS or VAMP-based algorithm.}

\begin{table}[htb]
	\color{black} 
	\centering
	\renewcommand\arraystretch{1.2}
	\caption{Complexity Comparison}\label{tab:Complexity}
	\begin{tabular}{>{\centering}p{40pt}>{\centering \arraybackslash }p{160pt}}
		\hline
		Algorithm & Complexity  \\
		\hline
		UAMP & $\mathcal{O}(NLKM)+ \mathcal{O}(NKM)$  \\	
		ALS & $\mathcal{O}(NLKM+L(M+K)N^{2})+ \mathcal{O}(N^{3})$  \\	
		VAMP & $\mathcal{O}((M+K)N^{3}+(M+K)N^{2})$  \\
		\hline					
	\end{tabular}
\end{table}

\begin{figure*}[t]
	\begin{small}
		\begin{align*} \label{bigP}   
			\boldsymbol{\mathcal{P}} =
			\begin{bmatrix}
				\mathbb{E}\left\{ \left( \frac{\partial f}{\partial h_{1,1}} \right)\! \left( \frac{\partial f}{\partial h_{1,1}^{*}}\right) \right\}&\ldots & \mathbb{E}\left\{\left(\frac{\partial f}{\partial h_{1,1}} \right)\left( \frac{\partial f}{\partial h_{K,N}^{*}}\right) \right\}& \mathbb{E}\left\{ \left( \frac{\partial f}{\partial h_{1,1}} \right)\! \left( \frac{\partial f}{\partial g_{1,1}^{*}}\right) \right\}&\ldots &\mathbb{E}\left\{ \left( \frac{\partial f}{\partial h_{1,1}} \right)\! \left( \frac{\partial f}{\partial g_{M,N}^{*}}\right) \right\}\\
				\vdots & &\vdots &\vdots& &\vdots\\
				\mathbb{E}\left\{ \left( \frac{\partial f}{\partial h_{K,N}} \right)\! \left( \frac{\partial f}{\partial h_{1,1}^{*}}\right) \right\} & \ldots & \mathbb{E}\left\{ \left( \frac{\partial f}{\partial h_{K,N}} \right)\! \left( \frac{\partial f}{\partial h_{K,N}^{*}}\right) \right\}& \mathbb{E}\left\{ \left( \frac{\partial f}{\partial h_{K,N}} \right)\! \left( \frac{\partial f}{\partial g_{1,1}^{*}}\right) \right\} &\ldots &\mathbb{E}\left\{ \left( \frac{\partial f}{\partial h_{K,N}} \right) \left( \frac{\partial f}{\partial g_{M,N}^{*}}\right) \right\}\\
				\\
				\mathbb{E}\left\{ \left( \frac{\partial f}{\partial g_{1,1}} \right)\! \left( \frac{\partial f}{\partial h_{1,1}^{*}}\right) \right\}&\ldots & \mathbb{E}\left\{\left(\frac{\partial f}{\partial g_{1,1}} \right)\left( \frac{\partial f}{\partial h_{K,N}^{*}}\right) \right\}& \mathbb{E}\left\{ \left( \frac{\partial f}{\partial g_{1,1}} \right)\! \left( \frac{\partial f}{\partial g_{1,1}^{*}}\right) \right\}&\ldots &\mathbb{E}\left\{ \left( \frac{\partial f}{\partial g_{1,1}} \right)\! \left( \frac{\partial f}{\partial g_{M,N}^{*}}\right) \right\}\\ 
				\vdots & &\vdots &\vdots& &\vdots \\ 
				\mathbb{E}\left\{ \left( \frac{\partial f}{\partial g_{M,N}} \right)\! \left( \frac{\partial f}{\partial h_{1,1}^{*}}\right) \right\} & \ldots & \mathbb{E}\left\{ \left( \frac{\partial f}{\partial g_{M,N}} \right)\! \left( \frac{\partial f}{\partial h_{K,N}^{*}}\right) \right\}& \mathbb{E}\left\{ \left( \frac{\partial f}{\partial g_{M,N}} \right)\! \left( \frac{\partial f}{\partial g_{1,1}^{*}}\right) \right\}& \ldots &\mathbb{E}\left\{ \left( \frac{\partial f}{\partial g_{M,N}} \right) \left( \frac{\partial f}{\partial g_{M,N}^{*}}\right) \right\}\\
			\end{bmatrix}.
				\tag{86} 
			\end{align*}
			\hrulefill
		\end{small}
	\end{figure*}

\section{Cramér-Rao Lower Bound}\label{sec:CRLB}
In this section, we derive the CRLB for the RIS channel estimation, which is used to serve as another performance benchmark, besides the ALS and VAMP-based algorithms.

We firstly rewrite the system model as
\begin{align}\label{rewrite_setY}
	\boldsymbol{\tilde{Y}}=(\boldsymbol{H}^T  \odot \boldsymbol{G})\boldsymbol{\Phi}^{T}+\boldsymbol{\tilde{W}},
\end{align}
%where $\boldsymbol{\tilde{Y}}=[\boldsymbol{\tilde{y}}_{1},\ldots,\boldsymbol{\tilde{y}}_{L}]$ and $\boldsymbol{W}=[\boldsymbol{w}_{1},\ldots,\boldsymbol{w}_{L}]$. Further, we have $\boldsymbol{\tilde{y}}_{l}=[\boldsymbol{y}^{T}_{1,l},\ldots,\boldsymbol{y}^{T}_{K,l}]^{T}\in \mathbb{C}^{L \times 1}$ and $\boldsymbol{w}_{l}=[\boldsymbol{w}^{T}_{1,l},\ldots,\boldsymbol{w}^{T}_{K,l}]^{T}\in \mathbb{C}^{L \times 1}$. 
and define a complex parameter $\boldsymbol{\theta}\in \mathbb{C}^{2N(M+K)\times 1}$, which includes all of unknown complex parameters in  $\boldsymbol{H}$ and $\boldsymbol{G}$ as
\begin{align}\label{theta}
	\boldsymbol{\theta} \triangleq\left[\boldsymbol{h}_{1}^{T},\ldots,\boldsymbol{h}_{N}^{T},\boldsymbol{g}_{1}^{T},\ldots,\boldsymbol{g}_{N}^{T},\boldsymbol{h}_{1}^{H},\ldots,\boldsymbol{h}_{N}^{H},\boldsymbol{g}_{1}^{H},\ldots,\boldsymbol{g}_{N}^{H} \right]^{T}. 
\end{align}
The likelihood function of {$\boldsymbol{\tilde{Y}}$} can be expressed as
{\begin{small}
	\begin{align}\label{likelihood}
		p(\boldsymbol{\tilde{Y}};\boldsymbol{\theta})\!=\!\left(\pi \sigma^{2}\right)^{-KML}\!\exp\! \left\{\!-{\sigma^{-2}} \sum_{l=1}^{L}\left\| \boldsymbol{\tilde{y}}_{l} \!-\!(\boldsymbol{H}^T  \! \odot \! \boldsymbol{G}) \boldsymbol{\Phi}^{T}_{l,:}\right\|^{2}\right\}\!,
	\end{align}
\end{small}}
and the logarithm of likelihood function can be expressed as
{\begin{small}
	\begin{align}\label{log_likelihood}
		\ln(p(\boldsymbol{\tilde{Y}};\boldsymbol{\theta}))\!\!=\!\!-KML \ln \left(\! \pi \sigma^{2}\right)\! \!-\!\!{\sigma^{-2}} \sum_{l=1}^{L}\left\|\boldsymbol{\tilde{y}}_{l}\!-\!(\boldsymbol{H}^T \!\! \odot \! \boldsymbol{G}) \boldsymbol{\Phi}^{T}_{l,:}\right\|^{2}\!\!,
	\end{align}
\end{small}}

Then, {define $f_{\boldsymbol{\theta}} \triangleq \ln(p(\boldsymbol{\tilde{Y}};\boldsymbol{\theta}))$ and the Fisher information matrix (FIM) $\boldsymbol{\mathcal{J}}_{\boldsymbol{\theta}}\in \mathbb{C}^{(2N(M+K))\times (2N(M+K))}$ can be obtained by
\begin{align}\label{Fisher}
	\boldsymbol{\mathcal{J}}_{\boldsymbol{\theta}}=\mathbb{E}\left\lbrace \left( \frac{\partial f_{\boldsymbol{\theta}} }{\partial \boldsymbol{\theta}}\right) \left( \frac{\partial f_{\boldsymbol{\theta}} }{\partial \boldsymbol{\theta}}\right)^{H}\right\rbrace .
\end{align} }
The partial derivatives of $f_{\boldsymbol{\theta}}$ with respect to $\boldsymbol{\theta}$ can be expressed as
\begin{small}
	\begin{align}
		\frac{\partial f_{\boldsymbol{\theta}}}{\partial h_{k,n}}\!
		=\! {\sigma^{-2}} \sum_{l=1}^{L} \!\left\lbrace [\boldsymbol{\Phi}^{T}_{l,:}]_{n} \boldsymbol{g}^{T}_{n} 
		\left (\!\!( [\boldsymbol{Y}^{'}_{l}]_{:,k}) )^{*}\!\!-\!((\boldsymbol{H}^{T})_{k,:}\!\odot\! \boldsymbol{G})^{*}(\boldsymbol{\Phi}^{T}_{l,:})^{*}\right )\!\right\rbrace,
		\label{h_mn_derivatives}
	\end{align}
\end{small}
\begin{small}
	\begin{align}
		\frac{\partial f_{\boldsymbol{\theta}}}{\partial g_{m,n}}\!=\! {\sigma^{-2}} \sum_{l=1}^{L} \left\lbrace [\boldsymbol{\Phi}^{T}_{l,:}]_{n} \boldsymbol{h}^{T}_{n} 
		\left ((\boldsymbol{Y}^{'}_{l})_{m,:})^{H}\!-\!(\boldsymbol{H}^{T} \odot  \boldsymbol{G}_{m,:})^{*}(\boldsymbol{\Phi}^{T}_{l,:})^{*}\right )\right\rbrace,
		\label{g_kn_derivatives}
	\end{align}
\end{small}
\begin{small}
	\begin{align}
	\frac{\partial f_{\boldsymbol{\theta}}}{\partial h_{k,n}^{*}}=\left( \frac{\partial f_{\boldsymbol{\theta}}}{\partial h_{k,n}}\right) ^{*},\qquad
		\frac{\partial f_{\boldsymbol{\theta}}}{\partial g_{m,n}^{*}}=\left( \frac{\partial f_{\boldsymbol{\theta}}}{\partial g_{m,n}}\right) ^{*}.
\end{align}
\end{small}where $ [\boldsymbol{\Phi}^{T}_{l,:}]_{n}$ is the $n$-th element of $\boldsymbol{\Phi}^{T}_{l,:}$, $\boldsymbol{Y}^{'}_{l}\triangleq[\boldsymbol{\tilde{y}}_{1,l},\ldots,\boldsymbol{\tilde{y}}_{K,l}]\in \mathbb{C}^{M \times K}$ and $\boldsymbol{\tilde{y}}_{k,l}\in \mathbb{C}^{M \times 1}$ represents the sub-vectors of $\boldsymbol{\tilde{y}}_{l}$ as 
	\begin{align}
\boldsymbol{\tilde{y}}_{l}=[\boldsymbol{\tilde{y}}_{1,l}^{T},\ldots,\boldsymbol{\tilde{y}}_{K,l}^{T}]^{T}.
    \end{align}
Hence, the FIM $\boldsymbol{\mathcal{J}}_{\boldsymbol{\theta}}$ can be expressed as
\begin{align}\label{reFisher}
	\boldsymbol{\mathcal{J}}_{\boldsymbol{\theta}}=\left[\begin{array}{cc}
		\boldsymbol{\mathcal{P}} &\boldsymbol{0} \\
		\boldsymbol{0} & \boldsymbol{\mathcal{P}}^{*}
	\end{array}\right],
\end{align}
where $\boldsymbol{\mathcal{P}}$ is shown in \eqref{bigP} at the top of next page and the size of $\boldsymbol{\mathcal{P}}$ and $	\boldsymbol{0}$ are $(N(M+K))\times(N(M+K))$.
The inverse of the FIM of $\boldsymbol{\theta}$ gives, under some regularity conditions, a lower bound for the augmented covariance matrix of an unbiased estimator of $\boldsymbol{\theta}$ as
%\begin{small}
	\begin{align}\label{J}
		\setcounter{equation}{86} 
		\boldsymbol{\mathcal{J}}_{\boldsymbol{\theta}}^{-1}=
		\left[\begin{array}{cc}
			\boldsymbol{\mathcal{P}}^{-1} &\boldsymbol{0} \\
			\boldsymbol{0} & (\boldsymbol{\mathcal{P}}^{-1})^{*}
		\end{array}\right],
	\end{align}
%\end{small}
where $\boldsymbol{\mathcal{P}}^{-1}$ is given by
%\begin{small}
	\begin{align}\label{Bound}
		\boldsymbol{\mathcal{P}}^{-1}=
		\left[\begin{array}{cc}
			\boldsymbol{\Omega}_{\boldsymbol{H}}&\boldsymbol{\mathcal{I}} \\
			\boldsymbol{\mathcal{I}}^{H} & \boldsymbol{\Omega}_{\boldsymbol{G}}
		\end{array}\right],
	\end{align}
%\end{small}
and $\boldsymbol{\Omega}_{\boldsymbol{H}}\in \mathbb{C}^{KN \times KN}$ and $\boldsymbol{\Omega}_{\boldsymbol{G}}\in \mathbb{C}^{MN \times MN}$ are the CRLB matrices for the estimates of $\boldsymbol{H}$ and $\boldsymbol{G}$, respectively, and $\boldsymbol{\mathcal{I}}\in \mathbb{C}^{KN \times MN}$ represents the remaining sub-matrices.

Furthermore, $\boldsymbol{\mathcal{P}}$ can be divided into four sub-matrices as
%\begin{small}
	\begin{align}\label{P}
		\boldsymbol{\mathcal{P}}=
		\left[\begin{array}{cc}
			\boldsymbol{\mathcal{P}}_{\boldsymbol{HH}} &\boldsymbol{\mathcal{P}}_{\boldsymbol{HG}}\\
			\boldsymbol{\mathcal{P}}_{\boldsymbol{HG}}^{H}& \boldsymbol{\mathcal{P}}_{\boldsymbol{GG}}
		\end{array}\right],
	\end{align}
%\end{small}
where $\boldsymbol{\mathcal{P}}_{\boldsymbol{HH}}\in \mathbb{C}^{KN \times KN}$, $\boldsymbol{\mathcal{P}}_{\boldsymbol{GG}}\in \mathbb{C}^{MN \times MN}$ and $\boldsymbol{\mathcal{P}}_{\boldsymbol{HG}}\in \mathbb{C}^{KN \times MN}$, as shown in \eqref{bigP}. According to the formula for inverse of a partitioned Hermitian matrix in \cite{2001bounds}, we can obtain $\boldsymbol{\Omega}_{\boldsymbol{H}}$ and $\boldsymbol{\Omega}_{\boldsymbol{G}}$ as following
\begin{align}
	\boldsymbol{\Omega}_{\boldsymbol{H}}=& (\boldsymbol{\mathcal{P}}_{\boldsymbol{HH}}-\boldsymbol{\mathcal{P}}_{\boldsymbol{HG}}\boldsymbol{\mathcal{P}}_{\boldsymbol{GG}}^{-1}\boldsymbol{\mathcal{P}}_{\boldsymbol{HG}}^{H})^{-1},\\
	\boldsymbol{\Omega}_{\boldsymbol{G}}=&(\boldsymbol{\mathcal{P}}_{\boldsymbol{GG}}-\boldsymbol{\mathcal{P}}_{\boldsymbol{HG}}^{H}\boldsymbol{\mathcal{P}}_{\boldsymbol{HH}}^{-1}\boldsymbol{\mathcal{P}}_{\boldsymbol{HG}})^{-1}.
\end{align}
So the CRLB of $\boldsymbol{H}$ and $\boldsymbol{G}$ can be donated as 
\begin{align}
	\mathrm{CRLB}_{\boldsymbol{H}}=\frac{\mathrm{trace}(\boldsymbol{\Omega}_{\boldsymbol{H}})}{KN},\\
	\mathrm{CRLB}_{\boldsymbol{G}}=\frac{\mathrm{trace}(\boldsymbol{\Omega}_{\boldsymbol{G}})}{MN}.
\end{align}

\section{Simulation Results}\label{sec:Simulation}

{In this section, we provide extensive numerical experiments to demonstrate the superior performance of the proposed UAMP-based channel estimation algorithm.} For comparison, we also include the ALS-based and VAMP-based channel estimation algorithms in \cite{2021Empowered}.
The threshold $\epsilon=10^{-3}$ and the maximum number of iterations is set to 30. The entries of $\boldsymbol{H}$ and $\boldsymbol{G}$ are independently drawn from {a complex Gaussian distribution with zero mean and unit variance. {The scaling ambiguity of the estimation is eliminated in the calculation of  the normalized mean square error (NMSE).
		%NMSEs combined with the true values of  ${\boldsymbol{H}}$ and ${\boldsymbol{G}}$.} 
		The SNR (in dB) is defined as}
	\begin{align}
		{\rm SNR}=&10\log_{10} \left(\frac{ N\mathbb{E}\left\lbrace \left \| \boldsymbol{\Phi}\right \|_{F}^2\right\rbrace }{L\beta^{-1}}\right).
	\end{align}
	
	{In order to estimate the RIS channels during the training phase, the RIS goes through $L$ configurations, which leads to the RIS phase matrix $\boldsymbol{\Phi}$ in \eqref{eq:Y_nopilot}. In the simulations, we consider two methods. One is that the RIS units are turned on or off randomly, leading to a matrix $\boldsymbol{\Phi}$ with entries 1 or 0 \cite{Mishra2019Channel}, which is called a binary matrix hereafter. In the simulations we assume that each entry in $\boldsymbol{\Phi}$ takes 1 or 0 with the same probability. The other one is that the phases of the RIS units are set to some discrete values, and in particular, the phase matrix $\boldsymbol{\Phi}$ is part of the DFT matrix (called partial DFT matrix), as in \cite{2019AnOptimal}.}
	
	We evaluate the performance of  estimators in terms of the NMSE of estimated channel matrices $\hat{\boldsymbol{H}}$ and $\hat{\boldsymbol{G}}$. 
	The NMSE performance of various estimators versus SNR with $L=N=K=M=64$ is shown in Fig. \ref{fig:NMSE_SNR_NLKM64}, {where Fig. \ref{fig:NMSE_SNR_NLKM64} (a) and (b) are for partial DFT matrix and binary phase matrix, respectively.} It is observed that the proposed UAMP-based algorithm significantly outperforms the ALS-based and VAMP-based channel estimators for both $\boldsymbol{H}$ and $\boldsymbol{G}$, {especially in the case that $\boldsymbol{\Phi}$ is a binary matrix.} %The results also indicate that DFT has the better performance than the sparse on/off phase matrix in RIS-aided channel estimation as shown in \cite{2019AnOptimal}. 
	Note that the NMSEs of $\boldsymbol{H}$ and $\boldsymbol{G}$ are very similar. To keep the figures clear, we only show the NMSE performance of $\boldsymbol{H}$ in the subsequent simulation results.
	
	\begin{figure}[t]
		\centering
		\includegraphics[width=0.8\columnwidth]{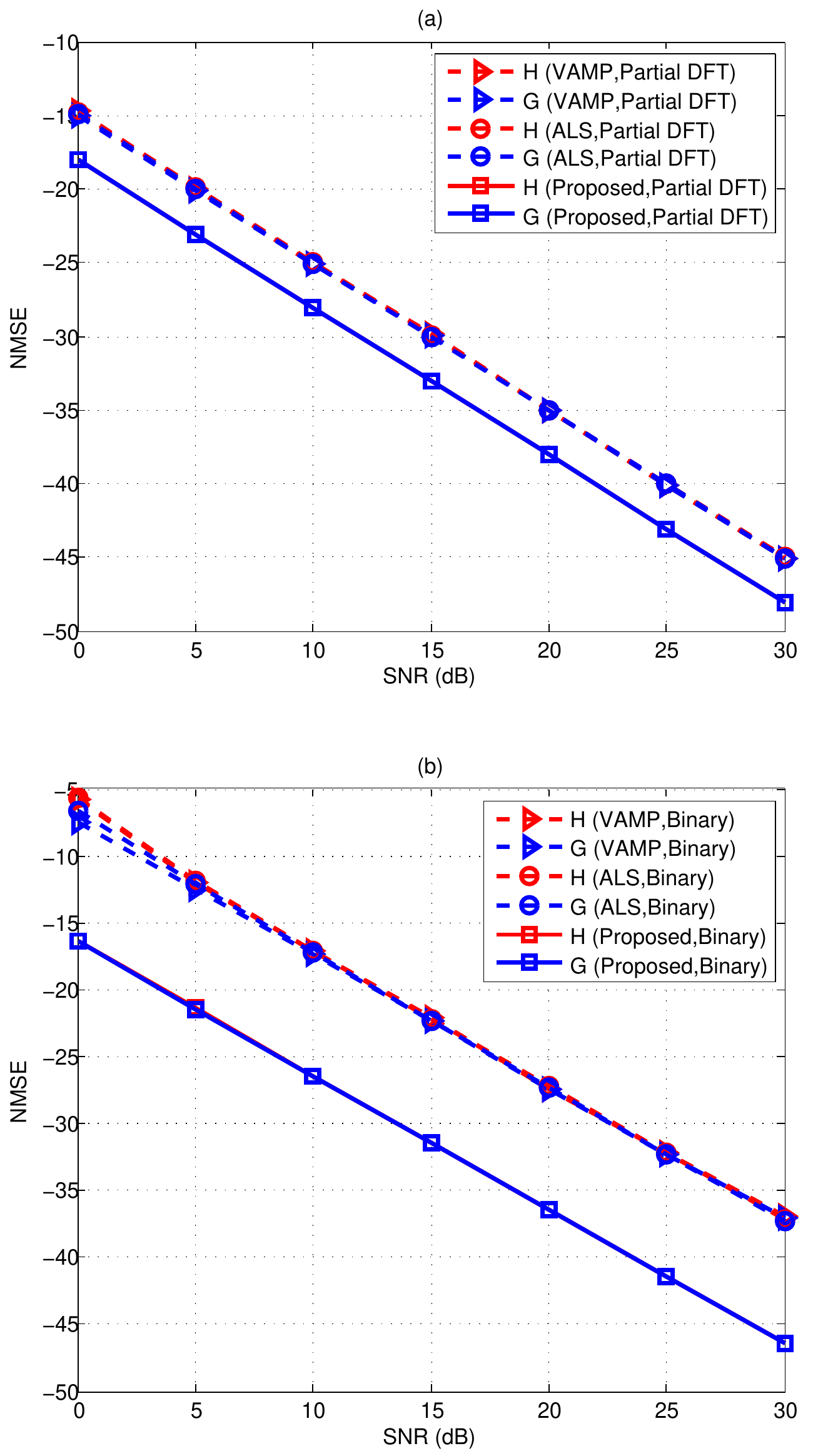}
		\caption{{ NMSE performance of the estimators versus SNR, where $L=N=K=M=64$. (a) Partial DFT matrix; (b) Binary matrix.}} \label{fig:NMSE_SNR_NLKM64}
	\end{figure}
	
	\begin{figure}[t]
		\centering
		\includegraphics[width=0.8\columnwidth]{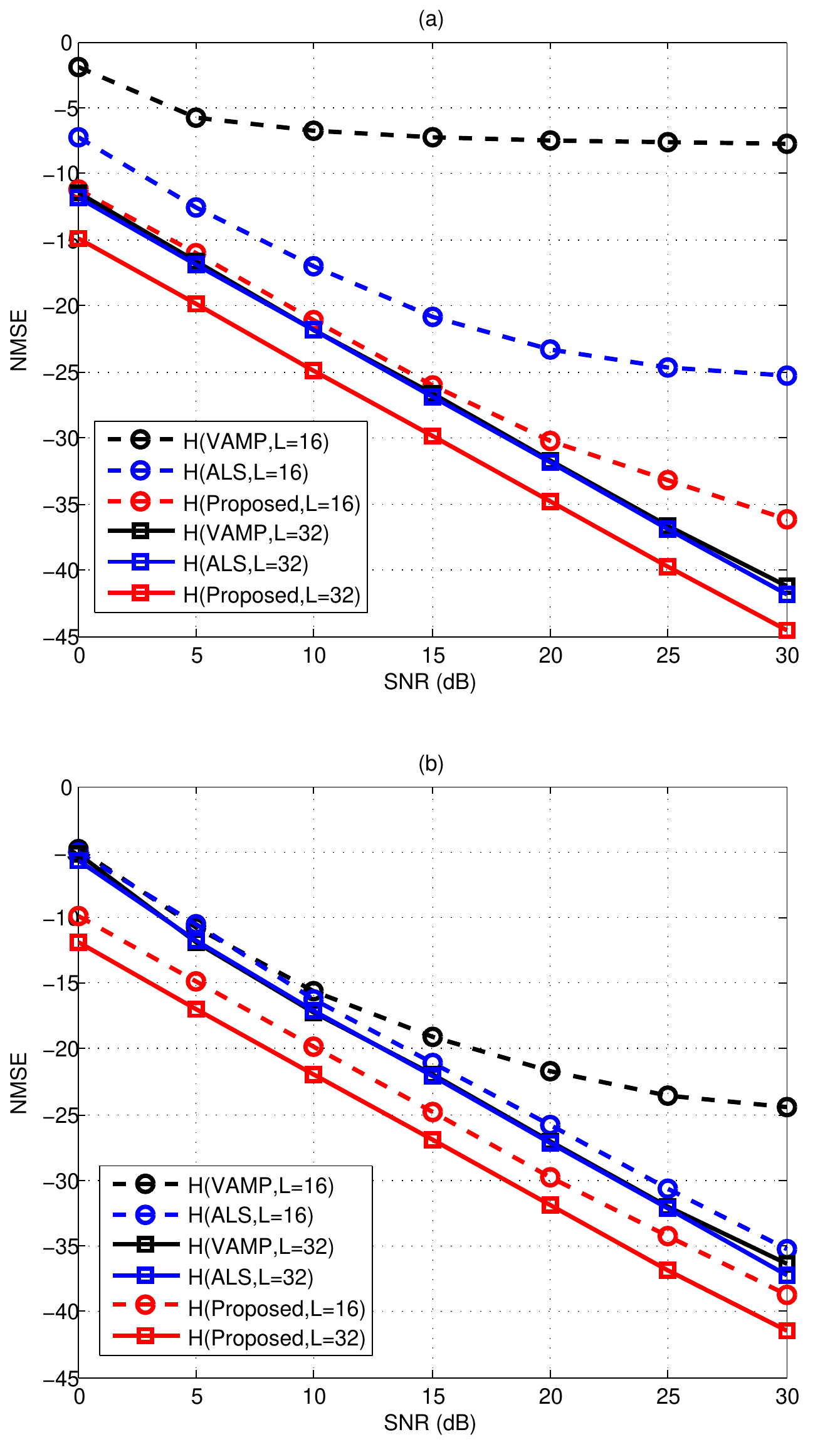}
		\caption{{NMSE performance of the estimators versus SNR for $N=K=M=64$ and different $L$ (a) Partial DFT matrix; (b) Binary matrix.}} \label{fig:H_var_L_SNR}
	\end{figure}
	
	\begin{figure}[t]
		\centering
		\includegraphics[width=0.8\columnwidth]{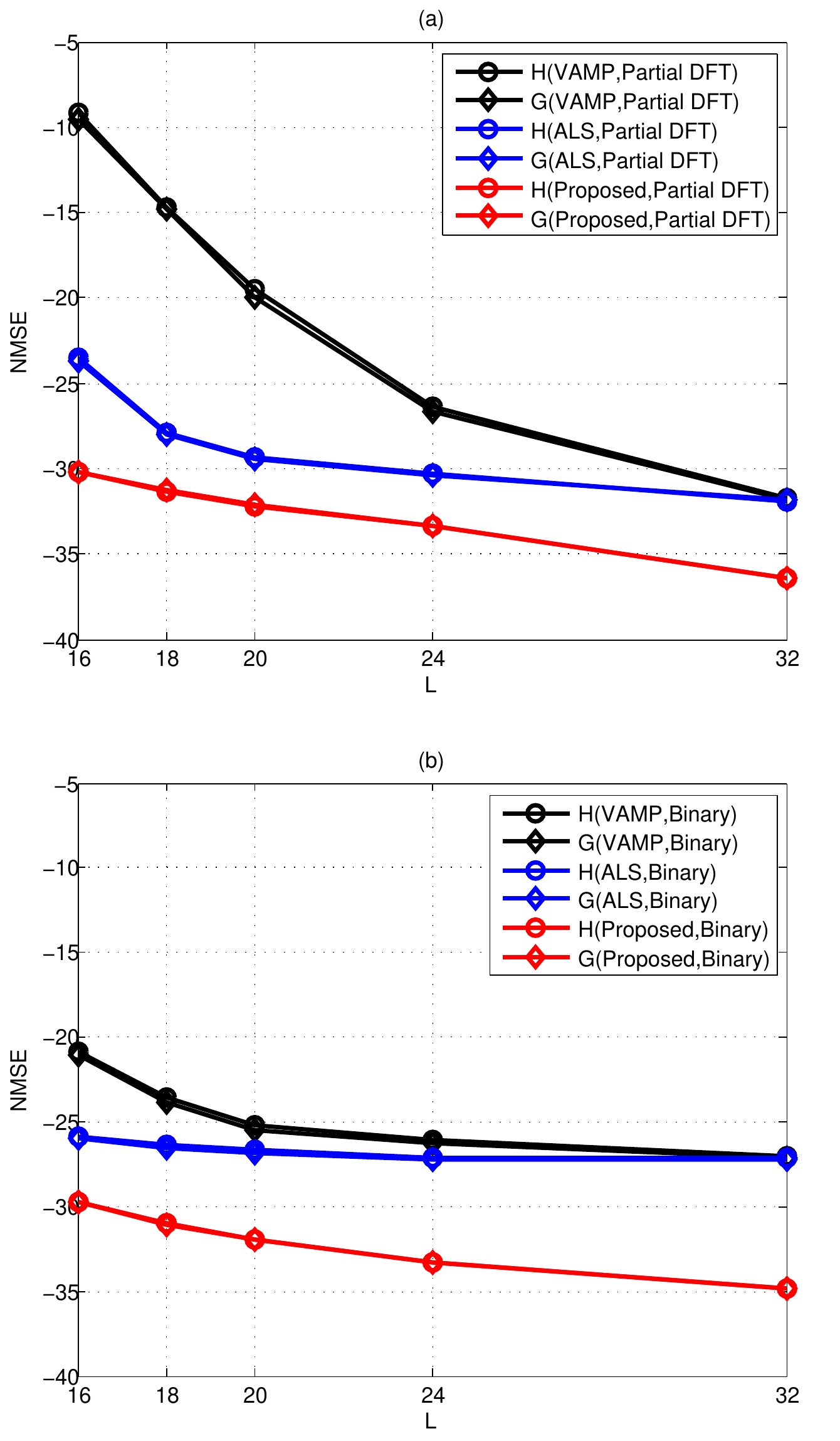}
		\caption{{NMSE performance of the estimators versus $L$, where $N=K=M=64$ and SNR=20dB. (a) {Partial} DFT matrix; (b) Binary matrix.}}\label{fig:var_L}
	\end{figure}
	
	{In Fig. \ref{fig:H_var_L_SNR}, with $N=K=M=64$, we compare the NMSE performance versus SNR of the estimators with different values of $L$. According to the results, the performance of the UAMP-based method is significantly better than that of the ALS and VAMP-based methods, especially when $L$ is relatively small.} {As smaller $L$ (the number of RIS phase configurations needed for channel estimation) is highly desirable to reduce the training overhead and latency,  next we vary the value of $L$ and examine the performance of the estimators. The results are shown in Fig. \ref{fig:var_L}, where the SNR is set to 20dB, and $\boldsymbol{\Phi}$ is partial DFT matrix in (a) and binary matrix in (b). It can be seen that, with the increase of $L$, the performance of all estimators improves, as expected. However, the performance improvement of the ALS and VAMP-based estimators is very slow with $L$. We can also see from Fig. \ref{fig:var_L} (a) that, the UAMP-based estimator with $L=16$ achieves the same performance of ALS and VAMP-based estimators with $L=32$. According to  Fig. \ref{fig:var_L} (b), the UAMP-based estimator with $L=16$ even outperforms the ALS and VAMP-based estimators with $L=32$. The results demonstrate that the use of the proposed algorithm can lead to a huge reduction in training overheads.}
	%The NMSE performances of algorithms for different $\boldsymbol{\Phi}$ and the number of available phase configurations $L$ are shown in Fig.\ref{fig:var_L}, which the SNR is set to 20dB. It can be seen that the increasing $L$ leads to a better performance as the training set is increasing in both figures, where the more training set is beneficial to the bilinear channel recovery problem.
	
	\begin{figure}[t]
		\centering
		\includegraphics[width=0.8\columnwidth]{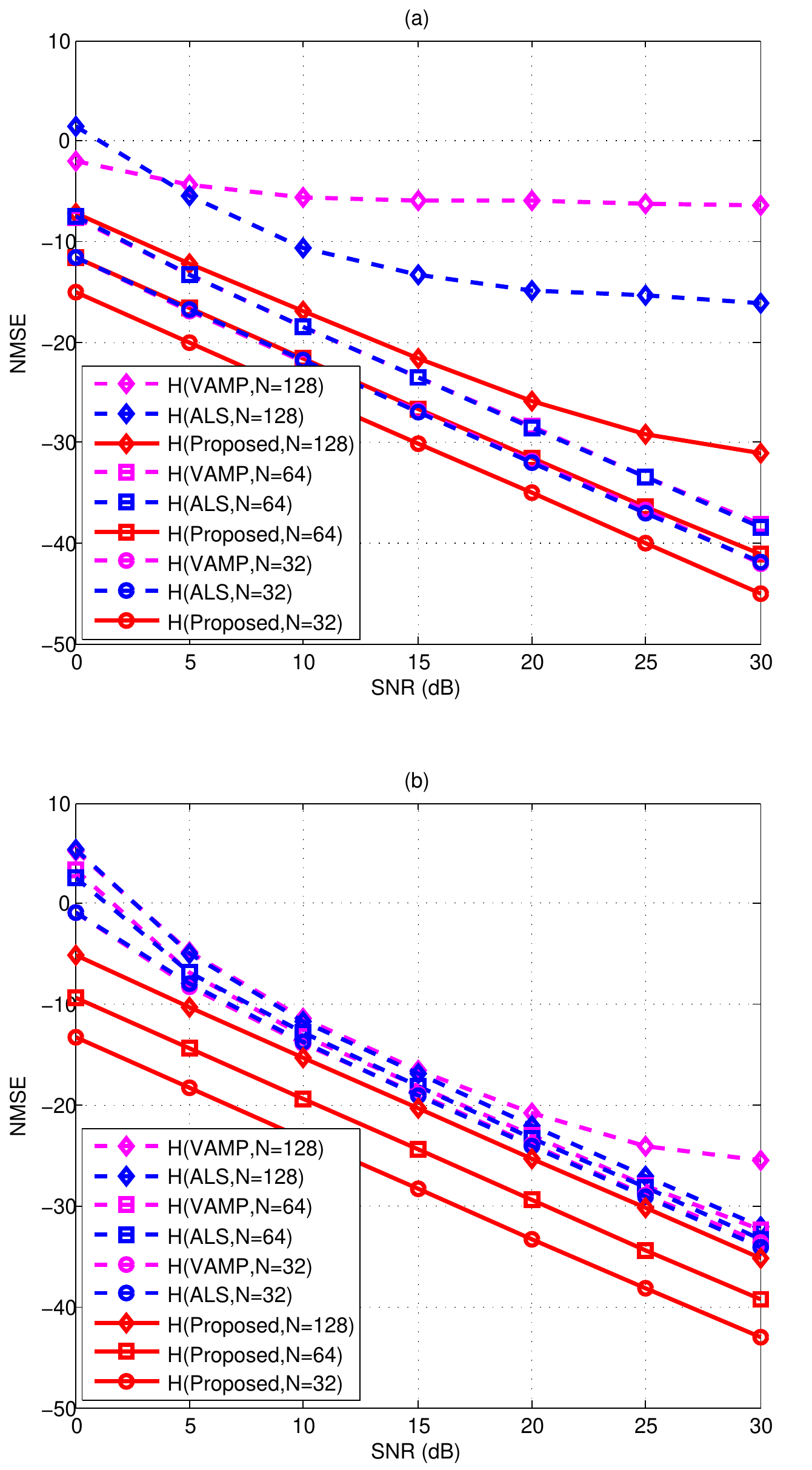}
		\caption{{NMSE performance of $\hat{\boldsymbol{H}}$ versus SNR with different $N$, where $L=K=M=32$. (a) Partial DFT matrix; (b) Binary matrix.}} \label{fig:var_N}
	\end{figure}
	
	\begin{figure}[t]
		\centering
		\includegraphics[width=0.8\columnwidth]{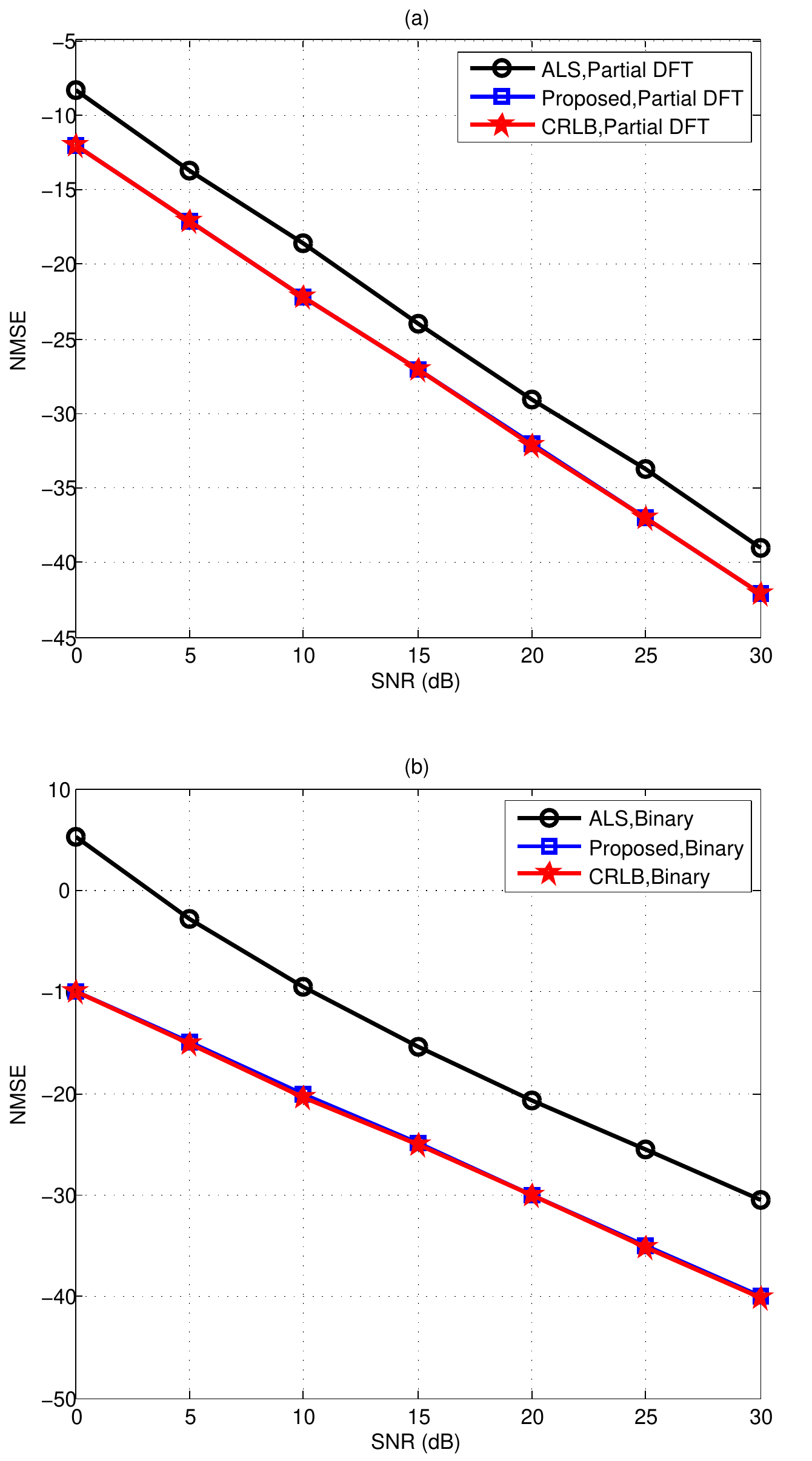}
		\caption{{Performance of the estimators and CRLB, where $L=K=M=N=16$. (a) Partial DFT matrix; (b) Binary matrix.}} \label{fig:CRLB_H}
	\end{figure}
	
	\begin{figure}[t]
		\centering
		\includegraphics[width=0.82\columnwidth]{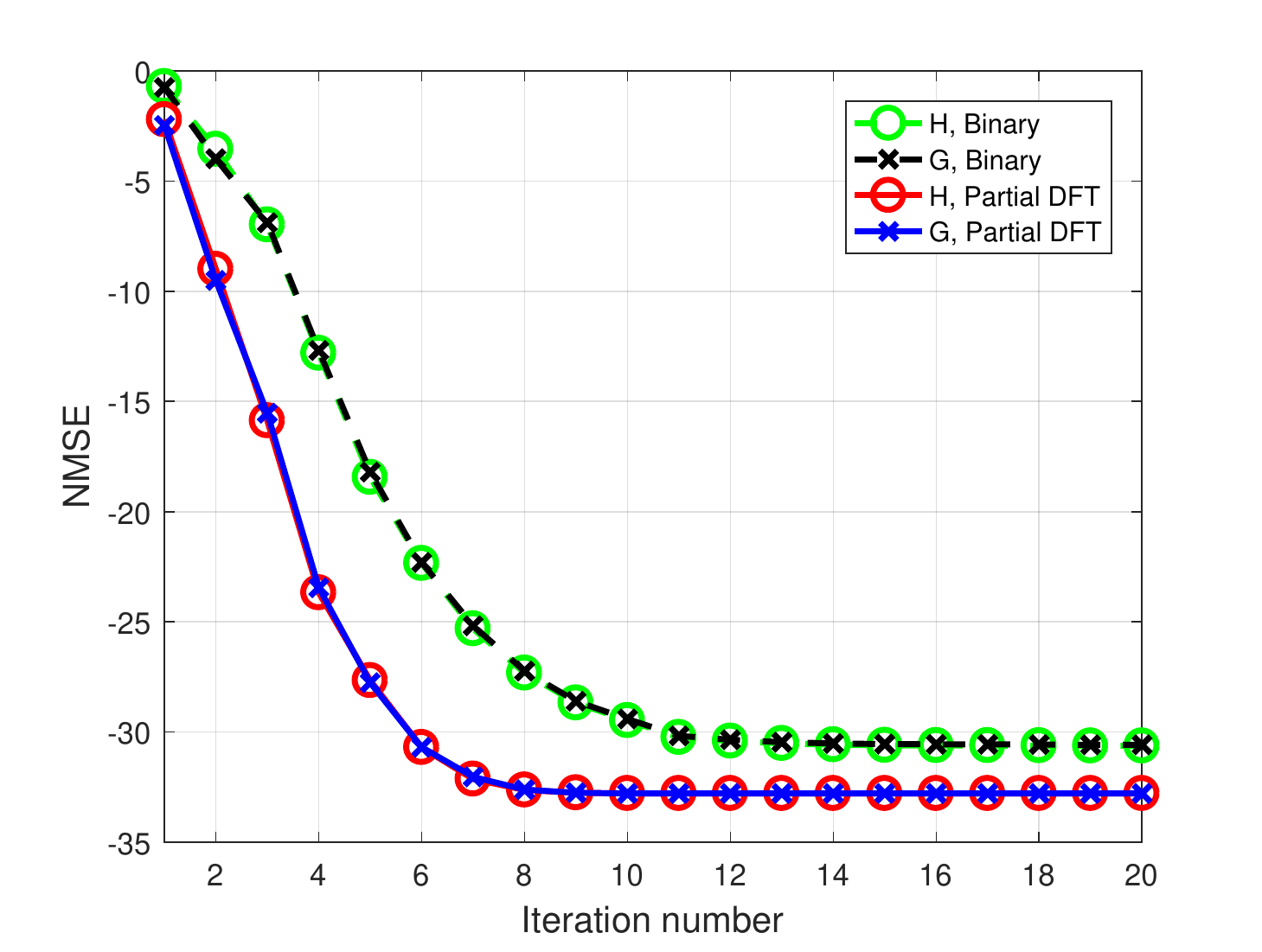}
		\caption{{NMSE performance of $\hat{\boldsymbol{H}}$ and  $\hat{\boldsymbol{G}}$ versus the iteration number, where $L=20$ and $K=M=N=32$.}} \label{fig:HG_iteration_vs_NMSE}
	\end{figure}
	
	\begin{figure}[t]
		\centering
		\includegraphics[width=0.82\columnwidth]{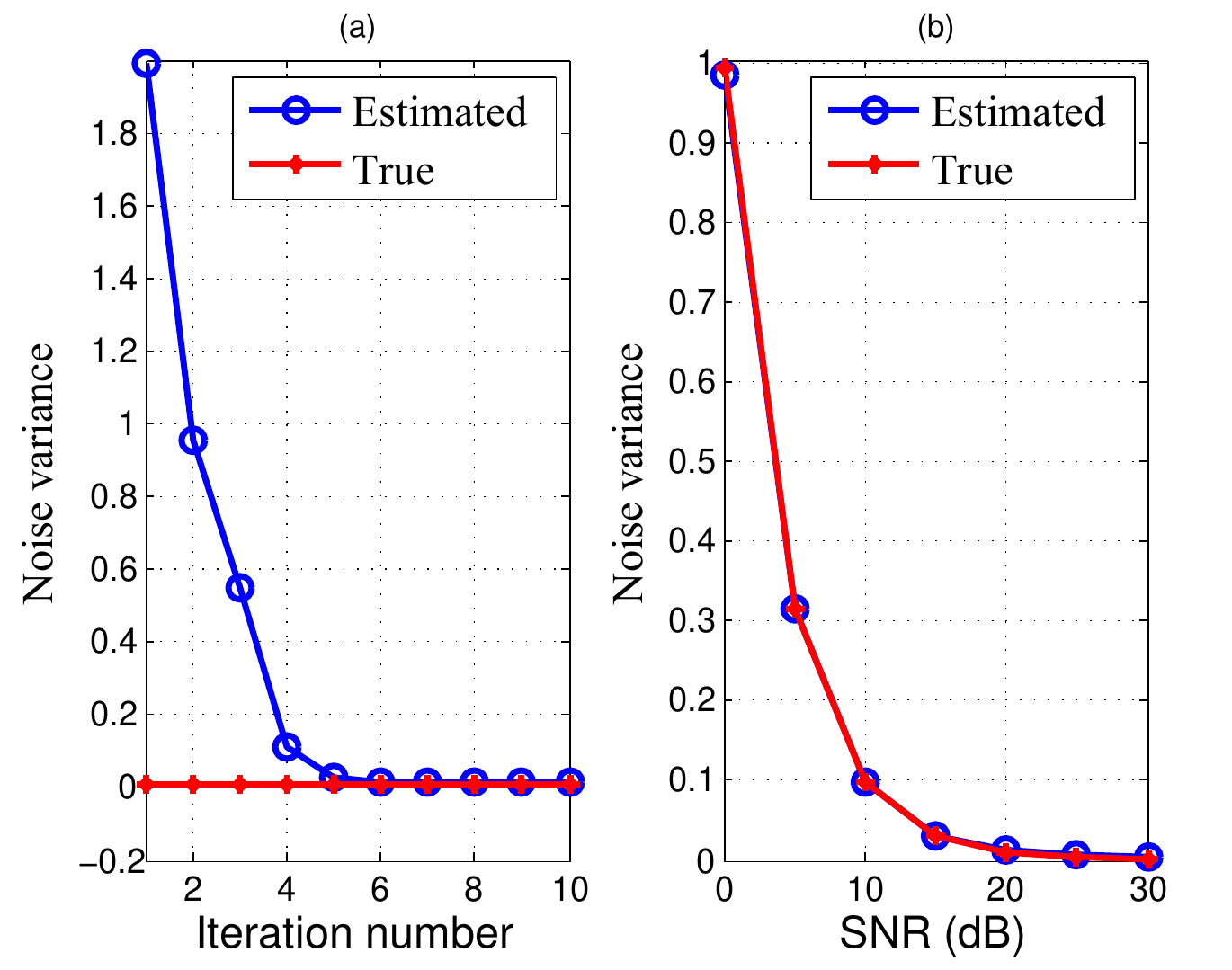}
		\caption{{Noise variance estimation of the proposed algorithm.}} \label{fig:precision}
	\end{figure}
	
	{We next examine the impact of the number of RIS units $N$ on the performance of channel estimation, where we set $L=K=M=32$. As a smaller $N$ leads to a less number of channel coefficients to be estimated, for a fixed $L$  the performance of the estimators improves with the decrease of $N$. In Fig. \ref{fig:var_N}, we can find that, in the case of binary matrix, the performance of the UAMP-based estimator with $N=128$ is even better than that of the ALS and VAMP-based estimators with $N=32$, which again demonstrates the superior performance of the proposed one. The performance of the UAMP-based algorithm, the ALS algorithm and the CRLB are shown in Fig. \ref{fig:CRLB_H}, where $L=K=M=N=16$. We can see that the performance of UAMP-based algorithm is almost the same as the CRLB, which is significantly better than that of the ALS algorithm. }
	
	{From the above results, we find that the performance of the proposed UAMP-based algorithm consistently show good performance for both partial DFT matrix and binary matrix. In contrast, the ALS and VAMP-based algorithms exhibit significantly worse performance in the case of binary matrix, compared to the partial DFT matrix.} 
	%As in the case of the binary matrix, the RIS units only need to be turned on and off randomly, it might be more convenient for training, compared to the partial DFT matrix. In this case, the proposed algorithm is more attractive.}
	
	{We also investigate the performance of the estimation with iteration number for different $\boldsymbol{\Phi}$ when the SNR is 20dB. The results are shown in Fig. \ref{fig:HG_iteration_vs_NMSE}. It can be seen that the proposed algorithm converges fast in different cases, especially when $\boldsymbol{\Phi}$ is a partial DFT matrix. It is worth mentioning that the estimation of the noise precision (reciprocal of the variance) is incorporated in the UAMP-based channel estimator, so no separate noise power estimator is needed. In Fig. \ref{fig:precision}, we compare the estimated noise power and its true value, where $\boldsymbol{\Phi}$ is a partial DFT matrix. In Fig. \ref{fig:precision} (a), the estimate of the noise variance with the iteration number is shown, where the SNR is 20dB. We can see that the convergence is fast. The results in Fig. \ref{fig:precision} (b) show that the proposed algorithm provides accurate noise variance estimates for a wide range of SNRs.}

\section{Conclusions} \label{sec:Conclusion}

{In this paper, we have addressed the issue of channel estimation in RIS-aided MIMO communications. Through vectorization and reduction, we obtain a new signal model for channel estimation, based on which a message passing based algorithm is developed, leveraging UAMP. Compared to the state-of-the-art algorithms, the proposed algorithm does not have any special requirements on the matrices involved, and it shows significant advantages in computational complexity, estimation performance and training overhead.
Extensive numerical results demonstrate the merits of the proposed algorithm.}

\bibliographystyle{IEEEtran}
\bibliography{IEEEabrv,bibliography}

\end{document}